\documentclass[10pt,journal,compsoc]{IEEEtran}

\ifCLASSOPTIONcompsoc
  \usepackage[nocompress]{cite}
\else
  \usepackage{cite}
\fi

\ifCLASSINFOpdf
   \usepackage[pdftex]{graphicx}
   \DeclareGraphicsExtensions{.pdf,.jpeg,.png}
\else
\fi
   
\usepackage{algorithm}
\usepackage{algorithmic}

\usepackage{tabularx}
\usepackage{tabulary}
\usepackage{array}

\ifCLASSOPTIONcompsoc
  \usepackage[caption=false,font=footnotesize,labelfont=sf,textfont=sf]{subfig}
\else
  \usepackage[caption=false,font=footnotesize]{subfig}
\fi

\begin{document}

\title{A System Architecture for the Detection of\\ Insider Attacks in Big Data Systems}

\author{Santosh~Aditham,~\IEEEmembership{Student Member,~IEEE}~and~Nagarajan~Ranganathan, ~\IEEEmembership{Fellow,~IEEE}
\IEEEcompsocitemizethanks{\IEEEcompsocthanksitem Both authors are with the Department
of Computer Science and Engineering, University of South Florida, Tampa, FL. USA.\protect\\
E-mail: saditham@mail.usf.edu}
}

\IEEEtitleabstractindextext{
\begin{abstract}
In big data systems, the infrastructure is such that large amounts of data are hosted away from the users. In such a system information security is considered as a major challenge. From a customer perspective, one of the big risks in adopting big data systems is in trusting the provider who designs and owns the infrastructure from accessing user data. Yet there does not exist much in the literature on detection of insider attacks. In this work, we propose a new system architecture in which insider attacks can be detected by utilizing the replication of data on various nodes in the system. The proposed system uses a two-step attack detection algorithm and a secure communication protocol to analyze processes executing in the system. The first step involves the construction of control instruction sequences for each process in the system. The second step involves the matching of these instruction sequences among the replica nodes. Initial experiments on real-world hadoop and spark tests show that the proposed system needs to consider only 20\% of the code to analyze a program and incurs 3.28\% time overhead. The proposed security system can be implemented and built for any big data system due to its extrinsic workflow.
\end{abstract}

}

\maketitle
\IEEEdisplaynontitleabstractindextext
\IEEEpeerreviewmaketitle
\IEEEraisesectionheading{\section{Introduction}\label{sec:introduction}}
\IEEEPARstart{B}{ig} data solutions are widely adopted across various government and enterprise domains such as software, finance, retail and healthcare. Big data applications are pioneering in the field of advanced data analytics and have a projected market of approximately 50 billion dollars by 2018. The most frequent use-cases of big data are information retrieval from complex, unstructured data; and real time data analysis \cite{IDC}. Along with its rapid market growth, the big data trend also has its share of challenges and risks. In an era where extracting information from data is sanctioned to all, users are understandably more skeptical to let providers host their data away from them. This, along with the recent increase in the number of cyber attacks, boosted the importance for security. Yet, the losses due to boundless security holes in existing systems seem to overshadow the investments towards increasing their security. Hence, there is an immediate need to address architectural loopholes in order to provide better security. For instance, current big data security platforms focus on providing fine-grained security through extensive analysis of stored data. But such models indirectly facilitate the abuse of user data in the hands of the provider.

Insider attacks are becoming more common and are considered the toughest attacks to detect \cite{Vormetric}. There does not exist much in the literature on solutions for insider attacks in general \cite{Salem}. Though privacy and security are touted to be important problems in the big data world, the solutions concentrate only on leveraging big data systems for efficient security in other domains. To the best of our knowledge, there is no robust solution for detecting or preventing insider threats within big data infrastructures. For example, security mechanisms of popular big data systems such as Hadoop \cite{Hadoop} and Spark \cite{Spark} include third-party applications such as Kerberos \cite{Kerberos}, access control lists (ACL), log monitoring and data encryption (to some extent). But for an insider, especially a traitor, circumventing these mechanisms is not difficult \cite{Aditham}. It is crucial to address the problem of insider attacks in big data systems for three main reasons: (a) traitor within the provider's organization will be able to circumvent the security system in place (b) sensitivity of customer information stored in the system is increasing by day; and (c) there is no consensus or widespread agreement on well-defined security standards in the big data community.

Recently, two unauthorized backdoors were discovered in Juniper Networks firewalls that might have given attackers access to highly classified information. Some important facts about this particular hack are: (a) it comes at the cost of compromising national security (b) it shows that even a major network security company is vulnerable to attacks (c) in spite of the high stakes and vast resources, it is believed that these backdoors were left undiscovered for almost 3 years; and (d) it was reported that the attackers could have deleted the security logs \cite{Swati}. This is one of the many examples to show that the efficiency of common attack prevention techniques, such as identity management, ACLs and data encryption, is necessary but sufficient to prevent attacks. As per OpenSOC, in 60\% of breaches data gets stolen within hours of the breach and 54\% of breaches are not discovered for months \cite{Sirota}. This indicates that infrastructures need to have efficient \textit{attack detection} techniques along with strong \textit{attack prevention} techniques for robust security.

In the big data world, it is considered that moving computation to where the data resides is better than the traditional approach of moving data for computation. The main features of big data infrastructures are fast data processing, high scalability, high availability and fault-tolerance. Availability and fault-tolerance of big data systems comes from intelligent replication of data. This implies SIMD style, parallel execution of the same program at multiple locations. When a program is scheduled for execution on the big data cluster, it runs as an individual process on every data node that hosts a copy of the program data. The replication of data on various nodes in the big data system can be utilized in providing security. Security for a computing system can be implemented at hardware and software level. Given the advantage of isolation that can be achieved at hardware level security, we propose delegating security to special purpose hardware, such as TPM \cite{TPM} and TXT \cite{TXT} chips, that reside on the nodes of the big data cluster. Such an infrastructure will have the advantages of (a) performing security analysis remotely (b) reducing the overhead on main processor by delegating security, and (c) significantly decrease the cost of data transfer while providing efficient security techniques such as isolated vulnerability scanning through program profiling.

In this paper, we propose a system architecture for attack detection in big data systems that can efficiently detect insider attacks. Our proposed system uses a two step algorithm for attack detection. First, \textit{program profiling} is performed by individual nodes of the big data cluster on the processes they execute. In this step, process binaries of scheduled processes are disassembled and analyzed to generate control instruction sequences (CIS). These sequences are then hashed, encrypted and shared among data nodes that host the same data i.e.\ primary and replica nodes. Next, \textit{consensus} among data nodes is achieved regarding the possibility of a process being attacked. This step involves two phases: hash matching and information sharing. Upon receiving encrypted messages from primary nodes, the replica nodes apply sequential, on-demand string matching between the locally generated hash and the received hash. Next, the result of this comparison is shared with the primary node. Depending on the results received, the primary data node notifies the master node to take necessary recovery measures. All communications among data nodes are performed using a \textit{secure communication protocol} that is based on public-private key encryption. Our main contributions are as follows: 
\begin{itemize}
\item We propose a novel extrinsic workflow for security in big data systems using control instruction sequences (CIS), hash matching and encrypted communication.
\item We suggest using a one-shot program profiling technique that builds instruction-level CIS from the native code of scheduled processes. 
\item We endorse the idea of having security as an independent module in big data systems by designing a system architecture for detecting insider attacks in big data systems. 
\end{itemize}

The paper is organized as follows: Section \ref{sec:background} gives a primer on insider attacks in general purpose computing. This section also discusses the current security methods in big data and some related works. Section \ref{sec:model} describes our attack model. The proposed system which includes the two step attack detection algorithm and the secure communication protocol are explained in Section \ref{sec:proposed}. A model of the proposed system with a list of required components is also given in this section. Section \ref{sec:experiments} shows the impact and usefulness of the proposed security system architecture by conducting real-world experiments on Hadoop and Spark clusters. Finally, section \ref{sec:conclusion} draws the conclusion and outlines future work.

\section{Background \& Related Work}\label{sec:background}
This section gives a primer on insider attacks and their solutions in general purpose computing and discusses the current security mechanisms available in the big data world. Also, the various related works are briefly described here.

\begin{figure}
\centering
\includegraphics[width=0.5\textwidth]{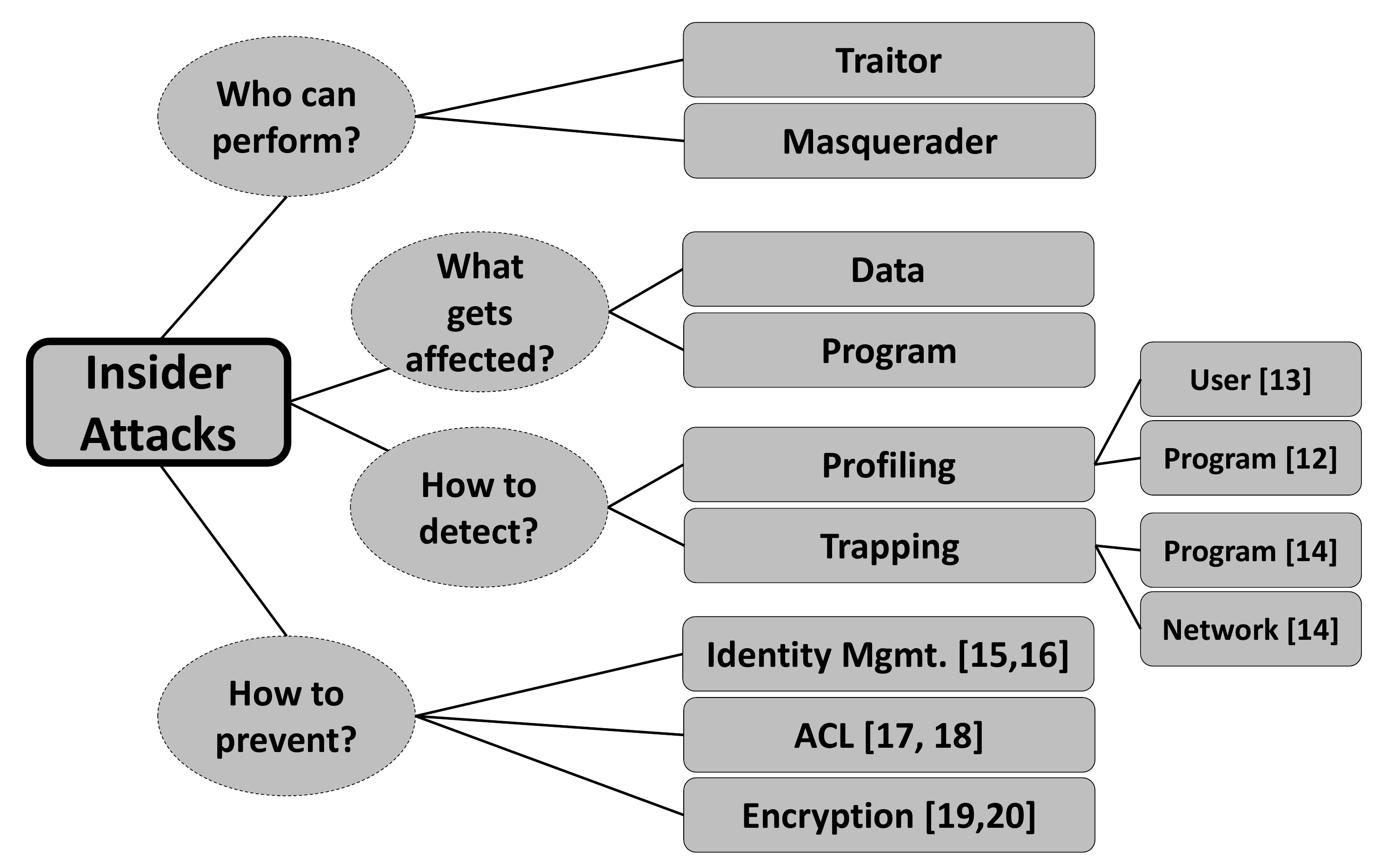}
\caption{Entities and Relationships in Insider Attacks}
\label{fig_ia}
\end{figure}

\begin{figure*}[!t]
\centering
\subfloat[Attack 1 - Manipulating Activity Logs]{ 
  \includegraphics[width=0.45\textwidth]{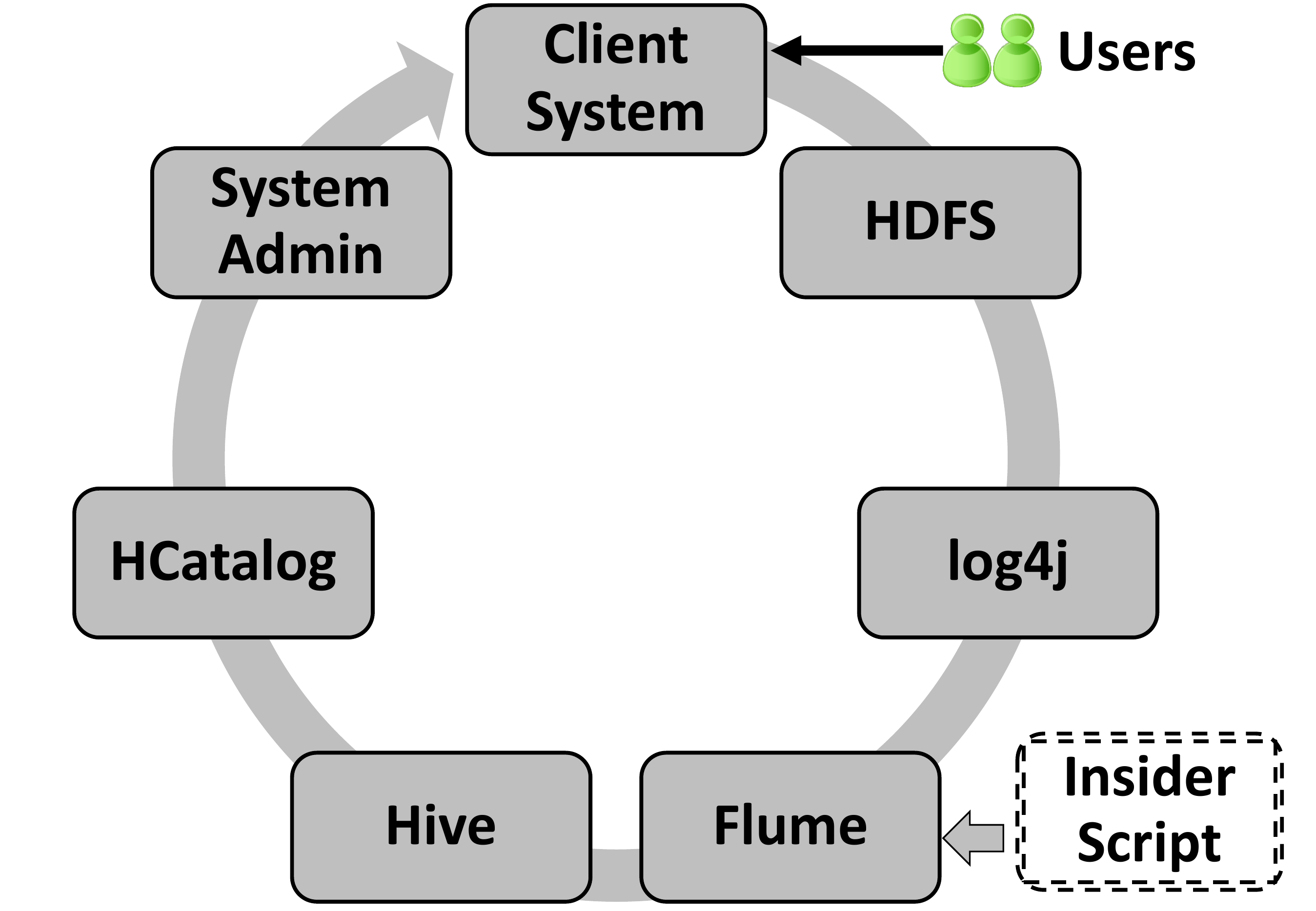}
  \label{fig_first_case}}
\hfil
\subfloat[Attack 2 - Deleting Edit Log]{
  \includegraphics[width=0.5\textwidth]{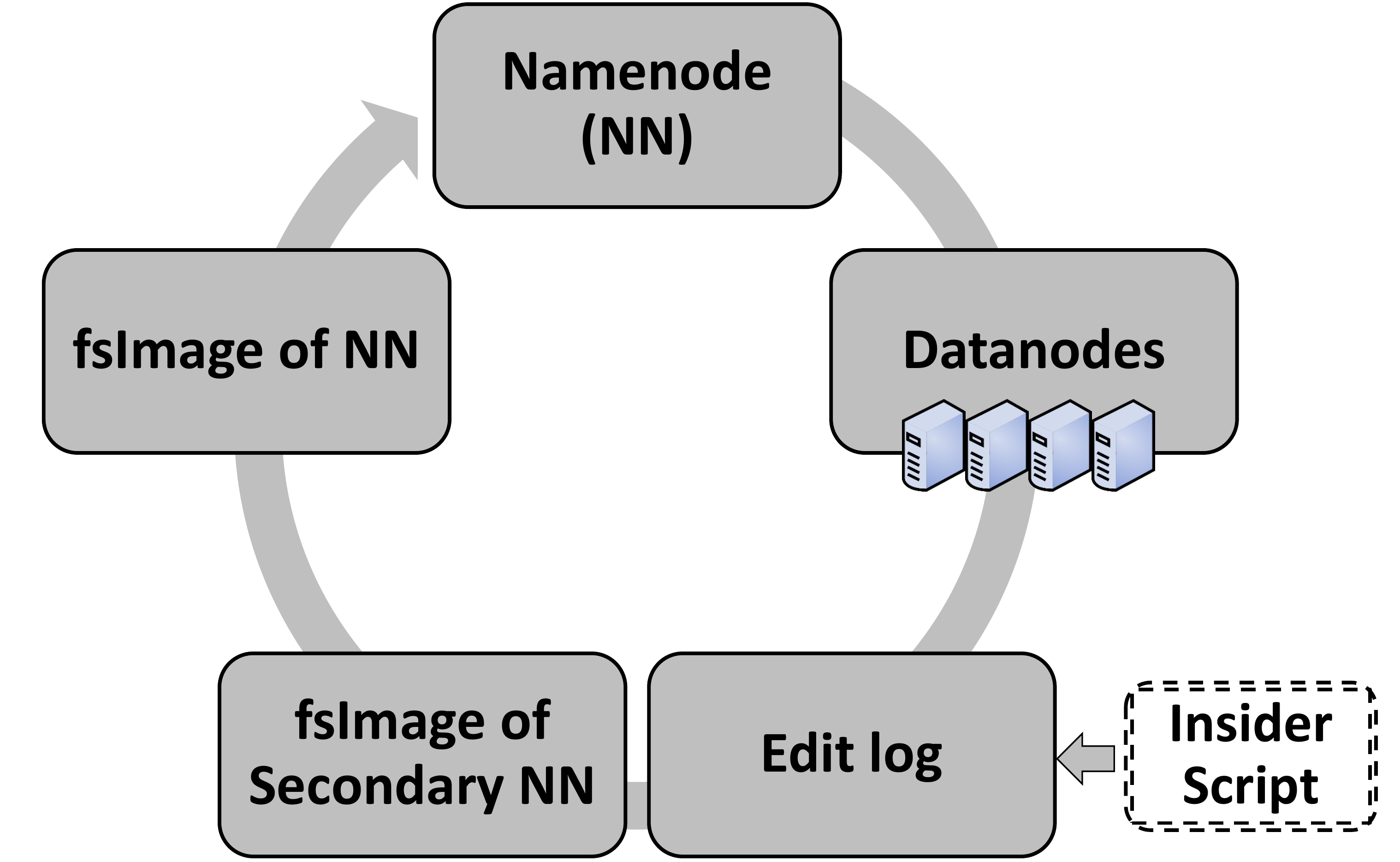}
  \label{fig_second_case}}
\caption{Examples of Security Loopholes in Hadoop}
\label{fig_bdsf}
\end{figure*}

\subsection{Insider Attacks}
Though security in general computing has been extensively studied and implemented over the years, computers are still vulnerable to attacks. Software based attacks that typically target a computer network or system, called cyberattacks, are growing in their frequency and impact. The plot for any type of software attack involves exploitation of a piece of code that runs on a computer. It is inherent to this perspective about a cyberattack that security can be provided at two levels: (a) by the software that is used to compile and execute the program; and (b) by the hardware that runs the program. Providing security at software level gives more context and information about the target programs that are being protected. But this comes with the risk of the security software itself being compromised. On the other hand, having security at hardware level gives more isolation to the process of analyzing and securing programs though it becomes difficult to give detailed context about the programs and the infrastructures running them. In any case, the toughest software attacks to counter are the ones whose genesis is intentional and are performed by those who have a good understanding of the underlying system.

Based on our literature review, we have identified four major questions that can guide towards better handling of insider attacks: (a) who can perform these attacks? (b) what gets affected? (c) how to detect these attacks? and (d) how to prevent them from happening? Figure \ref{fig_ia} gives a list of entities to consider when dealing with insider attacks. The figure also shows the four questions, from above, as relationships among the entities. Insider attacks can be performed by (a) \textit{traitors} who are legally a part of the system but want to misuse the access privileges given to them; (b) \textit{masqueraders} who get access to the system by stealing identities of those who have legitimate access. Insider attacks can affect the proper functionality of a program or corrupt the data used by the programs. Profiling and trapping are two most common ways to detect insider attacks \cite{Salem, Schultz}. Profiling can be performed (a) at the program level \cite{Anup} and at the user level \cite{Lunt}. Traps can be set in the programs or in the network to force the attacker into performing certain actions that help towards exposing the attack \cite{Spitzner}. The biggest concern with these insider attack detection methods is the possibility of losing valuable data. Hence, insider attack prevention mechanisms such as identity management \cite{Froomkin, Khalil2}, access control lists \cite{Ravi, Kerberos}, data encryption \cite{Don, Goyal} etc must be employed at the same time. 

In this work, we are more interested in Control Flow Integrity (CFI) \cite{Ligatti, Ligatti2} which is another popular and effective technique for attack prevention which enforces the execution of a program to follow a path that belongs to the program's control flow graph. The set of possible paths are determined ahead of time using static CFG \cite{Ligatti, Ligatti2}. A coarse-grained or fine-grained version of CFI can be used for program profiling. But the problem with any such profiling techniques is the overhead incurred in conducting them, even more if performed remotely. Though such limitations of this approach have been identified \cite{Jujutsu}, it is accepted as a strong and stable security enforcing mechanism. There are a plethora of CFG-based code similarity algorithms \cite{Chan}. But such CFG similarity check methods are complex, expensive, have no defined standards. Most CFG similarity algorithms rely on some simplification techniques such as fingerprints, edit distance, comparison only with known graphs in a database etc. Also, the impact of CFG similarity analysis differs a lot depending on when and how the CFG is generated for a program. These complexities and uncertainties led to a new set of control flow analysis techniques that avoid translating the program code to a formal model. For example, insider attack detection based on symbolic execution and model-checking of assembly code was proposed in \cite{Karthik}. In this work, we propose a novel approach for control flow similarity check for attack detection that totally discards the idea of building CFGs. Instead, our idea is based on simple string matching of control instruction sequences obtained from assembly code of scheduled processes. 

Insider attacks are a dangerous security problem in any domain because they are difficult to predict and detect \cite{Schultz}. Hence organizations must try to safe guard their systems and data from insider attacks \cite{Oltsik}. Predictive models for user/program/network behavior with the help of continuous monitoring is a widely adopted solution for insider attack detection. But such prediction is not completely reliable and the difficulty in detecting attacks grows with the complexity of the underlying system. Recent advancements in computing led to wide adoption of services such as cloud computing and big data which are extremely complex in their design and development. In cloud computing, many insider attacks can be performed by misleading the client side services and once compromised, data obtained can provide social engineering opportunities for cascade attacks \cite{Duncan}. Having security as a service model for cloud environments \cite{Vijay} and having sealed clouds \cite{Jager} are some ideas proposed towards protecting cloud infrastructures from insider attacks. While cloud computing is more about computing on the fly, big data deals with organizing and managing large sets of data. Insider attack detection and prevention for big data frameworks is an area that is not well explored yet. 

\subsection{Security in Big Data}
Security in big data is gaining tremendous momentum in both research and industry. But big data security is overwhelmingly inclined towards leveraging big data's potential in providing security for other systems \cite{Liang}. Security within big data systems is still a budding phenomenon. It is ideal to include security as a major component in the holistic view of big data systems. But the requirements of big data applications such as real-time data processing, fault tolerance and continuous availability give little scope to employ complex and robust security mechanisms. All existing security techniques implemented within big data frameworks are software based and try to prevent external entities from attacking the system. For example, the main requirements in hadoop security design focus only on access control \cite{Malley}. Big data systems encourage software based fine-grained security mechanisms such as Kerberos; access control lists (ACL); log monitoring etc. Big data security is inclined towards specifying multi-level access rights: user level, application level and data level. Advantages of having such simple software oriented security mechanisms, such as Kerberos, are better performance and simple management. But there are various problems with such a policy enforcing security software, as identified in \cite{Hu} and \cite{Gaddam}. Also, none of these approaches can strongly counter insider attacks.

According to Hadoop Security Design\cite{Malley}, permissible performance overhead for a change in architecture is only 3\%. This is precisely the reason behind coarse-grained security mechanisms such as data encryption being an optional and restricted feature in big data systems. Data encryption in hadoop is only available for data that gets exchanged between user and the system but not for data that travels within the system. Randomized data encryption for data security was proposed in \cite{Adluru} but this work acknowledges that faster results are yet to be achieved. Also, big data properties such as large scale distributed infrastructures and replication make it difficult to detect insider attacks precisely using the traditional methods. In this work, we demonstrate the inefficiency of existing big data security mechanisms by implementing two insider attacks on a big data cluster. Figure \ref{fig_bdsf} shows two workflows we used to successfully implement an insider attack in a hadoop big data cluster. This paper only discusses the workflow of the attacks but a detailed report on the results of these attacks can be found in our previous work \cite{Aditham}.

\subsubsection{Manipulating Activity Logs}
The first attack, as shown in Figure \ref{fig_first_case}, manipulates log data in order to produce erroneous results during log analysis. Flume and Kakfa are two popular big data products for real-time event processing. Most big data analysis and security solutions tend to use these services within their framework. Hortonworks tutorial \cite{Hortonworks} shows how a \textbf{system admin} will be able to detect distributed DOS attacks on the hadoop cluster by analyzing the server log data. Interestingly, this tutorial can be used as a counterexample to show that admin can act as a traitor, manipulate the server log data and create results that depict a wrong picture to the higher administration. As per the workflow in this example, users requests the client service to access data stored in HDFS. These user requests will all be logged by the log4j service. Hence, any attackers requests will also be logged. The system admin can easily build a framework with the help of services such as Flume, Hive and Hcatalog to monitor and track the user requests. A small script that filters the streaming data going from Flume to Hive can be induced by an insider to script the results according to the insider's choice. 

\subsubsection{Deleting Edit Log}
The second attack, as shown in Figure \ref{fig_second_case}, deletes the contents of \textit{editlog} such that user data gets deleted eventually. A \textbf{system admin} who has access to the \textit{secondary namenode} in a hadoop cluster can implement this attack. \textit{Namenode} is the focal point (and a single point of failure) of a HDFS file system. It stores the metadata of all files in HDFS along with their storage locations in a data blob called the \textit{fsImage}. Editlogs, along with fsImage, are updated periodically such that the namenode has access to up to date information about data stored in the hadoop cluster. To save time and computation energy on the namenode, this process is performed off-site on secondary namenode, sometimes called the \textit{checkpoint node} and the output fsImage is directly dumped on to the namenode. Hence, manipulating edit log content will reflect, by the next checkpoint, on the fsImage which will be used by the namenode for job allocation and scheduling. This is a weak point in the hadoop architecture that can be misused easily by insiders. Figure \ref{fig_second_case} shows the workflow for checkpoint-ing in a hadoop cluster and how an insider can introduce a script to delete user data forever. In the most extreme case, if an insider induces a small script that completely wipes out the editlog, the fsImage will be empty at the next checkpoint. 

Finally, proposing hardware oriented security methods for hadoop are on the rise in recent times. A TPM based authentication protocol for hadoop was proposed by \cite{Khalil} which claims to be much faster than Kerberos, though it has not been fully implemented. A hardware oriented security method to create trusted Apache Hadoop Distributed File System (HDFS) was proposed in \cite{Cohen} which is a theoretically novel concept but was proven to work only on one node. The overhead of data encryption by TPM acts as a hindrance in adopting this method, especially when the size of data maintained in big data systems is ever growing. In this work, we propose the delegation of security in big data systems by designing an independent system with the necessary components. 

\section{Attack Model}\label{sec:model}
Our attack model focuses on misuse of log data by system admins of a big data platform. Security features such as data confidentiality and operational guarantees such as correctness of results can be compromised because of such misuse. The goals of an insider conducting such attacks can vary from personal vendetta to financial gain. The proposed system targets such specific insider attacks because they are easy to implement with existing security solutions on platforms such as Hadoop and Spark. Attacks targeting misuse of log data can be performed by creating malicious programs or by modifying of existing program binaries with malicious intent. Given the existing security features of user-level activity monitoring, we exclude the possibility of system admins writing new malicious programs from the scope of our attack model. Instead, our attack model focuses on system admins being able to modify binaries of existing programs. Our goal is to spot  vulnerabilities in code that can be exploited by insiders. We acknowledge that insider attacks are too broad and not all of them can be mitigated by the proposed solution. There can be other possible insider attacks in big data that are not visible at compile time and the proposed system may or may not be able to detect. 

\section{Proposed System Architecture}\label{sec:proposed}
\begin{figure*}[!t]
\centering
\includegraphics[width=0.9\textwidth, height=3.5in]{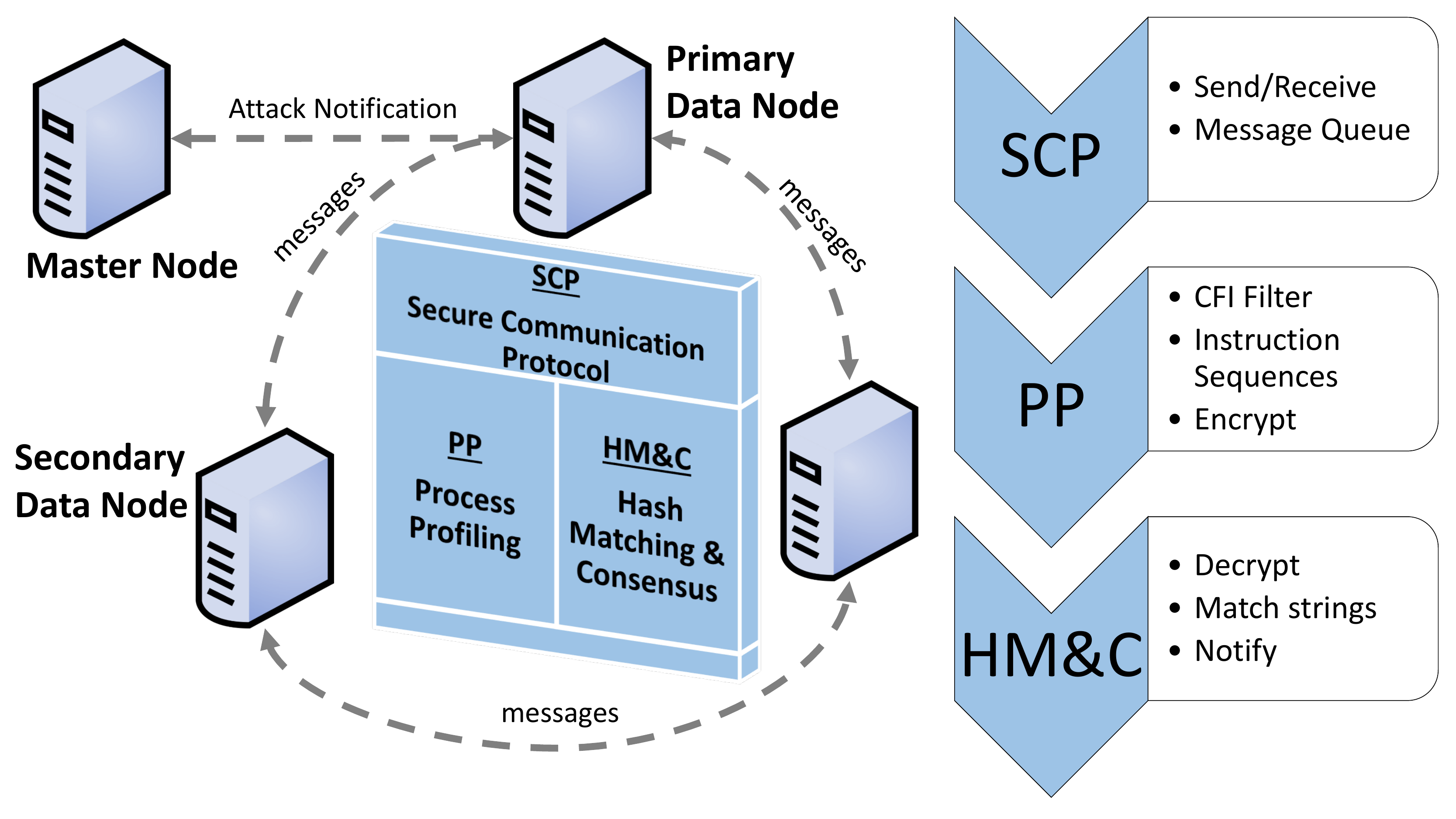}
\caption{Proposed System Architecture for Detecting Insider Attacks in Big Data Systems}
\label{fig_framework}
\end{figure*}

In this section we explain the proposed system in detail. Figure \ref{fig_framework} shows the proposed system that includes a secure communication protocol and a two step attack detection algorithm. The first step in the attack detection algorithm is process profiling, which is conducted locally and independently at each node to identify possible attacks. In the next step is hash matching and consensus, which is conducted by replica data nodes to conclude about the authenticity of a possible attack. 
 
\subsection{Secure Communication Protocol}

A big data system is technically a distributed data storage system that relies on secure and efficient communication protocols for data transfer. The proposed system aims to provide robust security for big data systems by having a modular design and being independent from the core big data services. For this reason, a separate secure communication protocol is included in the proposed system design that can be isolated from the set of default communication protocols used by the big data system. The proposed system is a mix of independent security modules that work together and reside on individual nodes of the system. These modules use the secure communication protocol to  share packets of data with their counterparts on other nodes of the cluster. 

The data shared among the security modules in our system architecture contain vital information about the analysis of a process. Hence, we propose using a public key cryptosystem in our secure communication protocol. All data transferred by any node using this secure communication channel is encrypted upfront using private key encryption and hardcoded keys that are not accessible to anyone. The associated public key will be shared with all other replica nodes that a data node need to communicate with. Hardware security chips such as TPM \cite{TPM} or Intel's TXT \cite{TXT} have public-private key encryption modules. Such hardware security chips come with a hardcoded, on-chip master key. A simple random number generator module is used to generate public-private key pairs periodically using the hardwired master key. For this work, we relied on SSH protocol for secure communication using RSA for key exchange but any such cryptosystem will work. Given the off chance of leakage of private keys, a key pair is held active for only a certain time period $T$. This increases the robustness of the communication protocol. In this work, we did not focus on finding the perfect value for $T$ but assumed it to be a predefined value of 1 second. The public key of a node is shared with all other nodes it has to communicate with i.e. replica nodes and master node. All incoming data packets to a node will be encrypted with its \textit{current} public key and can only be decrypted using the corresponding private key that is stored locally. Decrypted information will be sent to the \textit{process matching} module to identify attacks. 

Given the short lifespan of public keys used in our secure communication protocol, each node should be able to store public keys of all other nodes it has to communicate with. Also, storing older keys of other nodes helps in verifying authenticity of nodes in case of attack recovery. Hence, we propose to use queue data structures on every node to store the periodically generated public keys of other nodes. Back of $queue_{n}$ will be the latest public key to be used for encrypting packets to be sent to node $n$ while front of $queue_{n}$ will be deleted when $queue_{n}$ is full (to accommodate a new key). Limiting the maximum queue size by some $k$ will make sure that a node has enough information to support attack recovery measures while not consuming too much memory. Again, we did not focus on finding the perfect value for $k$ but used a predefined value of 3 while conducting our experiments.

Algorithm \ref{alg_sec_comm} shows the steps involved in the proposed secure communication protocol. Once a model of the proposed system is installed, all nodes will periodically generate public-private key pairs for as long as the system is in use. This is accomplished with the help of the hardwired key on the special purpose security chip and the random number generator module. At the end of every $T$ time units, a new public-private key ($newkp_n$) is generated on a node for communicating with replica node $n$. The private key $priv_n$ of $newkp_n$ will be used for decrypting incoming data from node $n$ and the public key $pub_n$ of $newkp_n$ will be shared with node $n$. For ease of access to keys during decryption, current private keys of all nodes are stored in an array $arr_{priv}[]$. Once a public key $pub_{n}$ is shared with node $n$, all incoming messages from node $n$ will only be decrypted using the associated $priv_{n}$ for the next $T$ time units. An array of queues, $arr_{pub}[]$, is used to store public keys received from all other nodes. When a node has to send an message $msg$ to replica nodes, the public key of that node is used to create an encrypted message $msg_e$.  

\begin{algorithm}[!t]
\caption{Secure Communication Protocol}
\label{alg_sec_comm}
\begin{algorithmic}
\WHILE{$true$}
    \IF {$time = T$}
    \FORALL{$n \in OtherNodes$}
      \STATE $newkp_{n} \gets$ get new public private key pair ($TPM$)
      \STATE $pub_{n} \gets$ get public key from $newkp_{n}$
      \STATE $priv_{n} \gets$ get private key from $newkp_{n}$
      \STATE $node_{n} \gets send (pub_{n})$
      \STATE $arr_{priv}[n] \gets priv_{n}$
    \ENDFOR
    \FORALL{$n \in OtherNodes$}
        \IF {$queue_{n}$ full}
           \STATE $dequeue(queue_{n})$
        \ENDIF
        \STATE $queue_{n} \gets enqueue (pub_{n})$
        \STATE $arr_{pub}[n] \gets queue_{n}$
    \ENDFOR
  \ENDIF
\ENDWHILE
\STATE $msg$ to be sent to all replicas
\FORALL{$r \in Replicas$}
  \STATE $pub_{r} \gets back(arr_{pub}[n])$
  \STATE $msg_{e} \gets encrypt(msg, pub_{r})$
  \STATE $send(msg_{e})$
\ENDFOR
\end{algorithmic}
\end{algorithm}

\subsection{Detection Algorithm}
The main part of the proposed system is the attack detection algorithm which will be explained in this subsection. Our attack detection algorithm is a two step process: process profiling (step 1) and consensus through hash matching (step 2).

\begin{algorithm}[!t]
\caption{Process Profiling}
\label{alg_profile}
\begin{algorithmic}
\WHILE{$true$}
  \STATE $proc_{new} \gets$ get newly scheduled process 
    \STATE $code \gets$ get assembly code from $HotSpotVM(proc_{new})$ 
  \FORALL{$instr \in code$}
    \IF {$instr \in jump$}
      \STATE $seq_{jump} \gets$ add $instr$ to sequence of jumps 
    \ENDIF
    \IF {$instr \in call$}
      \STATE $seq_{call} \gets$ add $instr$ to sequence of calls
    \ENDIF
    \IF {$instr \in return$}
      \STATE $seq_{return} \gets$ add $instr$ to sequence of returns
    \ENDIF
  \ENDFOR
  \STATE $seq_{array} \gets$ add $seq_{jump}$
  \STATE $seq_{array} \gets$ add $seq_{call}$
  \STATE $seq_{array} \gets$ add $seq_{return}$
  \FORALL{$seq \in seq_{array}$}
    \STATE $hash_{seq} \gets$ get hash from $sha(seq)$
    \STATE $hash_{hashes} \gets$ add $hash_{seq}$
  \ENDFOR
  \STATE $msg \gets$ get hash from $sha(hash_{hashes})$
  \STATE send $msg$ using Secure Communication Protocol
\ENDWHILE
 
\end{algorithmic}
\end{algorithm}

\subsubsection{Step 1: Process Profiling}
Traditionally vulnerability scanning is performed away from the source program's execution domain to guarantee isolation. Hence, the results of such scan must be communicated back to the program. But this leads to a cost versus isolation trade-off, depending on the remoteness of the location used to perform the vulnerability scan. In big data applications, the source program's execution is distributed across multiple nodes of the cluster. This makes it difficult to implement techniques such as vulnerability scans on big data systems. But big data infrastructures use replication of data for high availability. This enforces the same program to be run on multiple nodes that host the data required for the program. We exploit this unique property of big data systems and introduce a variation of CFI to create a novel process profiling technique that can help in detecting insider attacks in big data systems. Evans et al. \cite{Jujutsu} show that CFI, either with limited number of tags or unlimited number of tags, is not completely effective in attack prevention. Also, CFI is usually based on CFG created from static analysis of program code. 

Most big data applications are packaged as \textit{jars} that run on Java Virtual Machines (JVM). These jars are not completely compiled and do not convey much about the program they represent. Hence, we do not use CFI on CFG's created using statistical code analysis. We propose to build the control structure of a program from its corresponding JVM output i.e.\ the assembly code of the Hotspot VM that hosts the JVM. Since this is considered the final run-time code that gets executed on the hardware, the control structure generated from the output of Hotspot VM is expected to be less susceptible to software attacks compared to a CFG generated from statistical analysis of program code. In the context of big data platforms, this mitigates the possibility of launching an attack on the entire cluster. Another major variation from CFI in our process profiling technique is to use individual control flow instruction sequences instead of CFG paths. Control instructions dictate the control flow in a program. Generating instruction sequences of such control flow instructions from the assembly code output of hotspot VM should technically give us all information a CFG can provide in this context and avoid the complexity involved in generating a CFG. 

\begin{figure*}[!t]
\centering
\includegraphics[width=0.9\textwidth, height=3.25in]{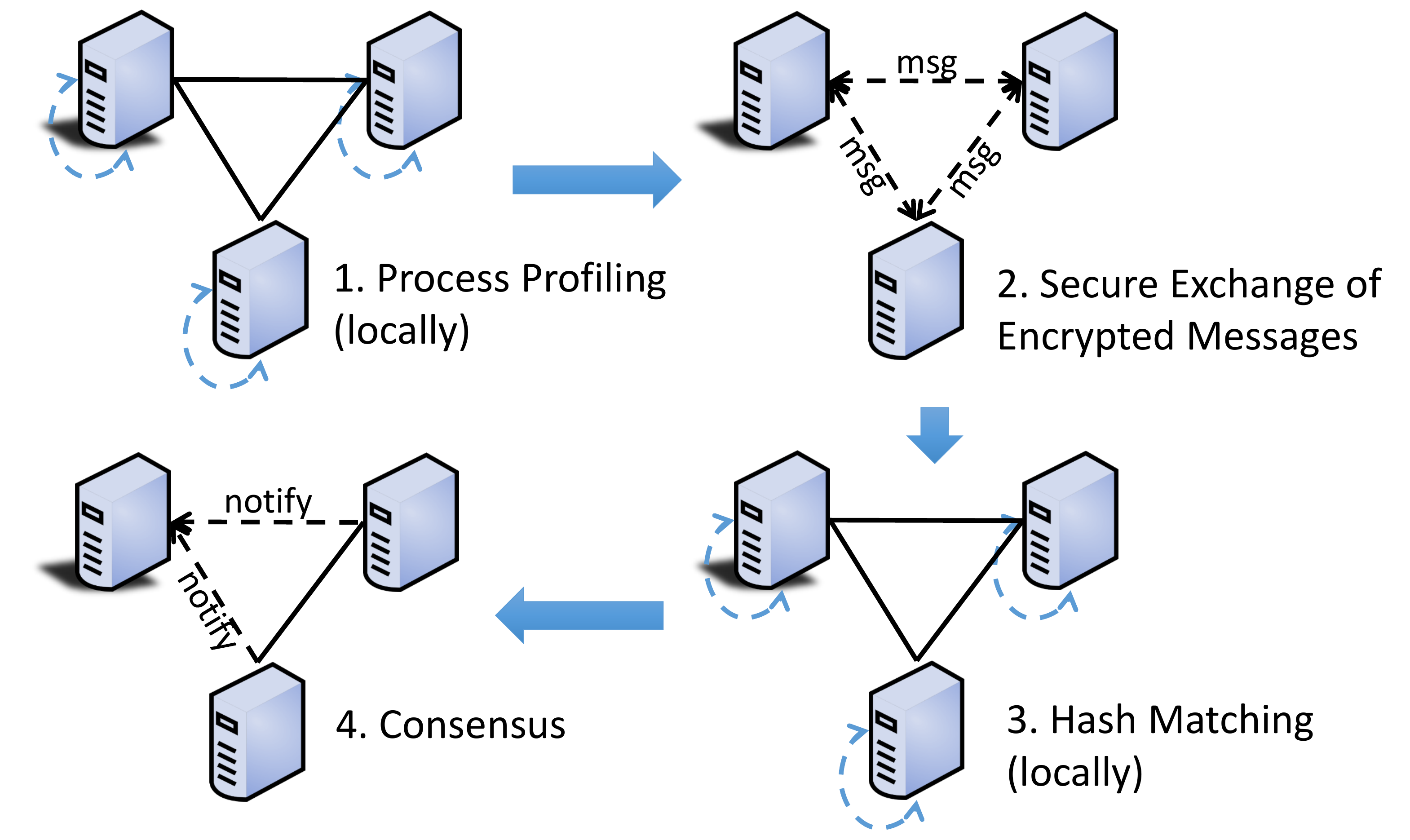}
\caption{Steps involved in the Detection of an Attack}
\label{fig_algo}
\end{figure*}

\subsubsection{Step 2: Hash Matching and Consensus}

\begin{algorithm}[!t]
\caption{Hash Matching}
\label{alg_match}
\begin{algorithmic}
\WHILE{$true$}
  \STATE $msg_{p} \gets$ get message about process p from main copy 
    \STATE $hash_{hashes}(received_{p}) \gets decrypt(msg_{new}, priv_{k})$ 
    \STATE $hash_{hashes}(local_{p}) \gets process-profile(p)$ 
  \IF {$hash_{hashes}(received_{p}) \in hash_{hashes}(local_{p})$}
      \STATE $confirmation \gets$ safe 
  \ELSE
      \STATE $confirmation \gets$ unsafe  
  \ENDIF
  \STATE $send(confirmation, main)$
\ENDWHILE
 
\end{algorithmic}
\end{algorithm}

The analyzer module in the proposed system creates instruction sequences for jumps, calls and returns from the JVM output of a given program (based on Intel's Instruction Set Architecture). Then, the SHA cryptographic hash function module is used to generate a fixed-length output for each of the three instruction sequences. All three hashes are combined and again given to the SHA cryptographic hash function module to generate a final hash for the program. This hash of hashes strengthens the uniqueness in identifying a program. All programs that run on every node in the cluster will follow the same routine.  Encryption module of the node with the primary copy of data uses currently active public keys of replica nodes to encrypt the hash of hashes and send it to the associated replica node. Hence, this node acts as the $coordinator$ for performing step 2 in the attack detection algorithm. 

Algorithm \ref{alg_profile} shows the steps involved in the proposed process profiling step. This algorithm will be running independently in the analyzer module of all machines in the big data cluster. Every process output, $proc_{new}$, from the HotSpot VM is grabbed by the analyzer module of the proposed system and profiled based on the control flow instructions present in its assembly code. Line by line analysis of $proc_{new}$ is conducted and each instruction $instr$ is matched with the set of control flow instructions available in the instruction set of the processor architecture. For this work, we used only the most prominent control flow instructions of Intel's x86 architecture i.e.\ jumps, calls and returns. When an $instr$ in the $code$ of the $proc_{new}$ is a control flow instruction, it gets added to the corresponding sequence string. The $seq_{array}$ represents the array of individual control flow instruction sequences in the process $proc_{new}$. This array is used later as input while generating the hashes for each control sequence string. All fixed length hash outputs are combined as $hash_{hashes}$ and rehashed to generate a final hash called $msg$ that represents the program. This $msg$ is then shared with all replicas running the same program using the secure communication protocol described above. 

The second step in our attack detection algorithm is a consensus algorithm similar to the extended 2-phase commit protocol \cite{Roger}. In this step, the node with primary copy of data acts as coordinator and requests all replica nodes, that act as workers, to confirm if their local hash of hashes, ($msg$) of a particular process matches exactly with the coordinator's version. The coordinator then decides on the safety of the process depending on the acknowledgments received from participating replica nodes. A process is considered to be safe by the coordinator if and only if it receives safe acknowledgments from all of the workers. At the end of process profiling step, encrypted message $msg_e$ is shared by coordinator node with all worker nodes. The nodes that receive such messages will decrypt the message  with their currently active private key. The decrypted message is essentially the hash of hashes of the three control instruction sequence strings. This decrypted hash of hashes can be directly compared to the local version of the same process to detect the possibility of an attack. If the result of such comparison of strings is a perfect match, then that indicates that the same process (with the same code) was run on both nodes. This indicates a safe process unless both nodes of the cluster are attacked the same way, in which case it will be a false positive. A confirmation message about the result of the hash comparison will be sent to the coordinator node as response to the original incoming message. The coordinator node will wait to receive responses from all replicas in order to arrive at a conclusion about the possibility of an attack in a process. The given big data system is safe as long as all the replicas respond with a \textit{safe} confirmation. A single \textit{unsafe} response will mean that the system is under attack. Algorithms \ref{alg_match} and \ref{alg_consensus} give more details about the hash matching and consensus steps that take place in step 2 of the attack detection algorithm. A pictorial representation of the steps involved in our 2-step attack detection algorithm is given in Figure \ref{fig_algo}. This figure represents a big data system with a replication factor of 3 and hence there is one coordinator (represented with a dark black shadow below the node) and two workers. Active communication channels are represented using a dotted line while the regular lines between nodes represent passive communication channel. The blue dotted loop around each node in step 1 and 3 of the figure represent local computations.

Algorithm \ref{alg_match} is used in the hash matching step of the attack detection algorithm. When a worker node, $node_k$ receives $msg_{p}$ from the coordinator node about a process $p$, it will decrypt that message using its current private key, $priv_{k}$ and stores the result as $hash_{hashes}(received_{p})$. The local version of the same string i.e.\ $hash_{hashes}(local_{p})$ will be compared against the $hash_{hashes}(received_{p})$ to identify similarity between local and received hash of a process. The result of this hash matching is sent back as $confirmation$ to the coordinator node, $main$. The value of $confirmation$ is \textit{safe} in case of a perfect match of hashes and \textit{unsafe} otherwise. 

\begin{algorithm}[!t]
\caption{Consensus}
\label{alg_consensus}
\begin{algorithmic}
\FORALL{$node \in replicas$}
  \STATE $confirmation_{node} \gets$ get confirmation about process $p$ from replica 
  \IF {$confirmation_{node} \in safe$}
      \STATE $count_{safe} \gets count_{safe} + 1 $ 
    \ENDIF
\ENDFOR
\IF {$count_{safe} \in count_{replicas}$}
  \STATE $attack \gets$ no 
\ELSE
  \STATE $attack \gets$ yes   
  \STATE $master_{node} \gets recovery(p)$
\ENDIF
 
\end{algorithmic}
\end{algorithm}   

Algorithm \ref{alg_consensus} is used by the coordinator node to identify an attack, with the help of worker nodes. After step 1, the coordinator node waits for responses from all the recipients. The worker nodes respond with a confirmation message that says whether the process is $safe$ or $unsafe$. If the count of number of $safe$ responses i.e.\ $count_{safe}$ from worker nodes matches with the count of number of $nodes$ in the replica set i.e.\ $count_{replicas}$, the coordinator node assumes that there is no attack in the current process $p$ and resets the $attack$ variable. Else, if a mismatch in the process analysis is observed, the $attack$ variable is set and the $master_{node}$ is notified about the possibility of an attack in process $p$.

\subsection{Model of the Proposed System Architecture}

The proposed security system is a combination of 3 parts: secure communication protocol, process profiling and hash matching. As shown in Figure \ref{fig_framework}, these three parts are made of multiple modules that need to be installed on all nodes in the big data system. Also, locality of these modules impacts the performance of the system greatly. The closer they are to the main processor of a node, the faster and less expensive it will be to communicate. But from a security standpoint, these modules need to be isolated from the big data system main workflow. Hence we designed a model for the proposed system that can fit on isolated special purpose security hardware chips. Such chips can be built on top of existing security hardware such as TPM or Intel's TXT chips \cite{TPM, TXT}. Hardware solutions are popularly known to affect the scalability and flexibility of the big data infrastructure, comparing to a software solution which can be very adaptive. But in this case, we avoid such problems by decoupling our solution from the workflow of a big data platform. There will be a one-time extra cost due to the hardware security modules. An overview of the elements in such a model of the proposed system is given in Figure \ref{fig_hardarch}. The functionality of each of these elements is as follows:

\begin{figure}
\centering
\includegraphics[width=0.45\textwidth]{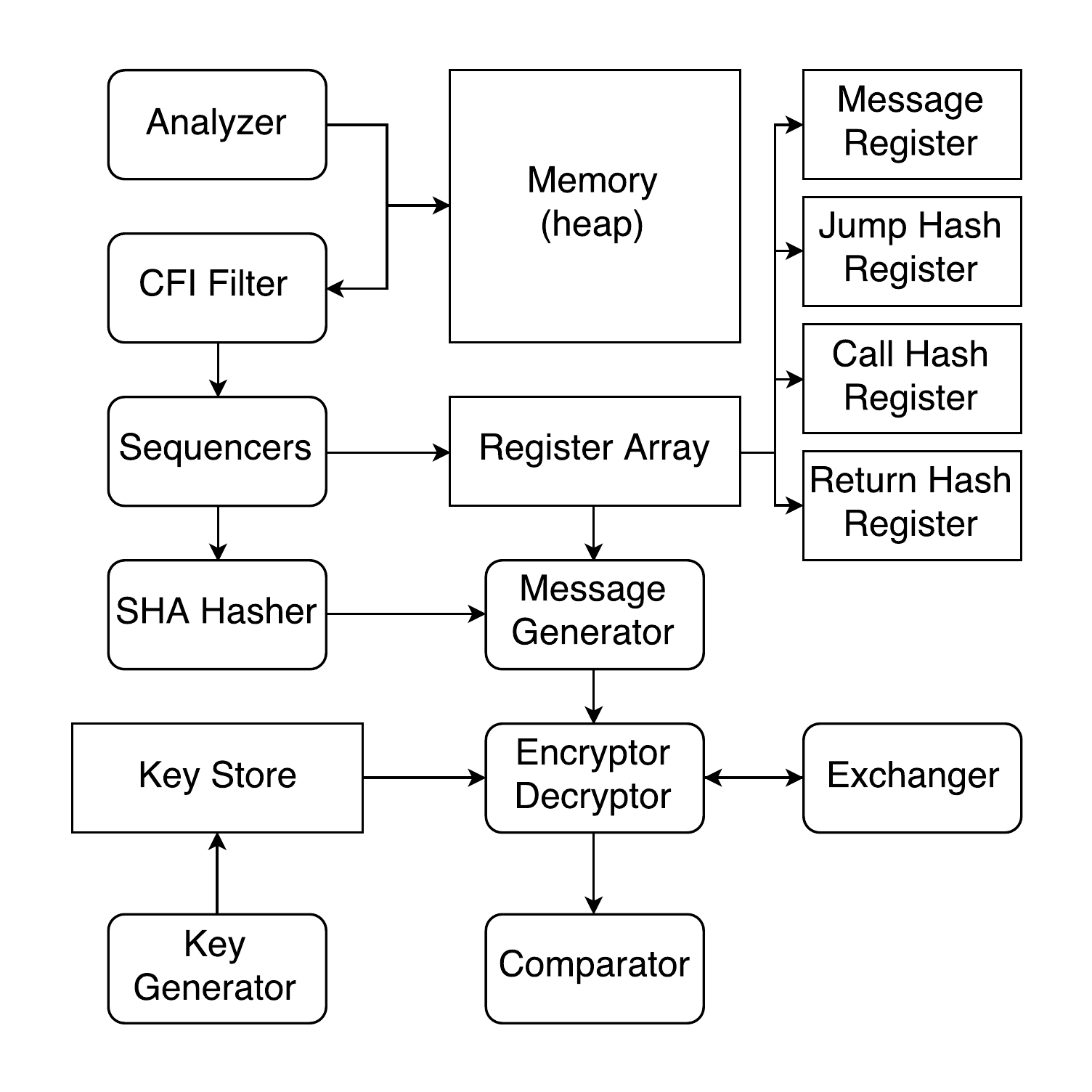}
\caption{Elements in a Model of the Proposed System Architecture}
\label{fig_hardarch}
\end{figure}

\begin{itemize}
  \item \textbf{Analyzer}, this module will get the data from the hotspot VM and perform the initial steps of cleaning the data. Result from analyzer is stored in \textit{Memory}.  
  \item \textbf{CFI filter}, this module takes input, a set of assembly language instructions, from the \textit{Analyzer} module (technically, the \textit{Memory} module) and filters out the control flow instructions, while maintaining the order.
  \item \textbf{Sequencers}, there are three sequencers in our model, one each for jumps, calls and returns. Each sequencer goes through the output of \textit{CFI filter} module and forms a delimited sequence string of the instruction it is associated with. Then, the sequencer uses the \textit{SHA hasher} module to generate and store a fixed length hash output from the variable length instruction sequence string. 
  \item \textbf{Register Array}, there are 4 registers in this array to store message, jump instruction hash, call instruction hash and return instruction hash.
  \item \textbf{Message Register}, this is a special register in the \textit{Register Array} used to store the message in thread-safe manner.
  \item \textbf{Message Generator}, this module combines all the individual hash outputs stored in registers and uses the \textit{SHA hasher} module to generate a fixed length hash output. This hash of hashes is combined with the process metadata to generate and store a message that represents the process.
  \item \textbf{Encryptor / Decryptor}, this module uses the \textit{Key Store} to access the current set of public/private keys and the \textit{Message Register} to access the current process message. The Encryptor module uses the public key of a replica node from the \textit{Key Store} and encrypts the message in \textit{Message Register}. The decryptor module uses the private key of the node from the \textit{Key Store} to decrypt an incoming message.
  \item \textbf{Comparator}, this module performs string comparison between local message (hash of hashes) and received message.
  \item \textbf{Key Generator}, this module uses the underlying TPM/TXT chip's \cite{TPM, TXT} in-built functionality. The hardwired key and the random number generator of the security chip are used to generate a new public/private key pair; and the timer of the chip to trigger this action periodically. 
  \item \textbf{Key Store}, this module uses an array of memory locations to store the public key queues of all replica nodes and the current public / private key pair of this node. The three most recent public keys of each replica node is stored in its queue.
  \item \textbf{Exchanger}, this module uses TCP/IP protocol to exchange messages with other nodes.
\end{itemize}

\section{Experiments and Results}\label{sec:experiments}
In this section we describe the experimental setup, explain in detail about our choice of experiments and analyze the results. The hadoop security design specifies that a 3\% slowdown in performance is permissible for any newly proposed security solutions \cite{Malley}. Hence, it is important for the proposed system to offer both theoretical correctness and feasibility in practical implementation and usage. Security in big data systems is a new area that does not have set standards and specifically designed open-source benchmarks to evaluate the overhead. Hence, we had to handpick a set of general big data benchmark programs that are relevant and provided by the big data community to test the efficiency of our proposed security system. 

\subsection{Setup}
The 3 big data services used for our experiments are:
\begin{itemize}
\item \textbf{Hadoop \cite{Hadoop}}, the most popular implementation of a big data framework that is maintained by the Apache open-source community. It allows storing and processing of large date using programming models such as MapReduce. 
\item \textbf{Spark \cite{Spark}}, a fast and general engine for large-scale data processing that is supposedly much faster than Hadoop and it is maintained by the Apache open-source community as well.  
\item \textbf{Amazon web services (AWS) \cite{AWS, EC2, EBS}}, a perfect example of real-world big data system. AWS provides Elastic Cloud Compute (EC2) service that allows users to use Amazon cloud's compute capacity depending on their needs. EC2 presents a true virtual computing environment. Storage for the EC2 nodes is provided by Amazon Elastic Block Store (EBS) which offers persistent storage. EBS volumes are automatically replicated to protect user from component failure, offering high availability and durability.
\end{itemize}

\begin{table}[!t]
  \caption{Amazon EC2 instance types}
  \label{table_ec2}
  \centering
  \begin{tabulary}{0.5\textwidth}{|C|C|C|}
    \hline
    \textbf{Attribute} & \multicolumn{2}{c|}{\textbf{Value}} \\
    \hline
    Instance Model & t2.micro & m1.large \\
    \hline
    Processor & Intel Xeon with Turbo & Intel Xeon E5-2650 \\
    \hline
    Compute Units & 1 (Burstable) & 4\\
    \hline
    vCPU & 1 & 2 \\
    \hline
    Memory (GB) & 1 & 7.5 \\
    \hline
    Storage (SSD) & Elastic Block Store & Elastic Block Store \\
    \hline
    Networking Performance & low & moderate \\
    \hline
    Operating System & Linux/UNIX & Linux/UNIX \\
    \hline
    Hadoop distribution &2.7.1 & 2.7.1\\
    \hline
    Spark distribution &N/A & 1.6\\
    \hline
  \end{tabulary}
\end{table}

We used AWS supported hadoop and spark clusters for conducting our experiments. The \textit{Hadoop Cluster} that we used is a 5 node cluster built using basic t2.micro nodes of Amazon EC2 and EBS. Each node is equipped with only 1 vCPU and 1GB memory. The network performance is minimal for this cluster. The \textit{Spark Cluster} that we used is a 4 node cluster built using general purpose m1.large nodes of Amazon EC2 and EBS. Each node is equipped with 2 vCPU and 7.5GB memory. Network performance is moderate for this cluster. Both cluster configurations satisfy the minimum requirement to support replication factor of 3. The hardware and software configurations of the EC2 nodes can be found in table \ref{table_ec2}. We built a 64-bit Ubuntu AMI (Amazon Machine Instance) for each node-type before setting up the clusters. These AMIs were equipped with the latest distributions of Hadoop, Spark and GCC along with with our code base. The hadoop cluster had 5 nodes, where 1 node acted as the namenode, 1 node acted as the secondary name node and 3 nodes were acting as data nodes. The spark cluster had a master and 3 slave nodes. Since our proposed system works independently, all modules of the model had to be installed on every node of the EC2 clusters. A library of all modules in the model was implemented in C++ programming language using STL and multi-threading libraries and packaged together. Our code used TCP/IP protocol and SSH keys for communication between the nodes of the clusters.

\begin{table}[!t]
      \caption{List of Hadoop Map Reduce Examples}
      \label{table_examples_hadoop}
      \centering
      \begin{tabulary}{0.45\textwidth}{| C | C | C |} \hline
            \textbf{Exp.no}&\textbf{Name}&\textbf{Description} \\ \hline
            1&  aggregatewordcount  &An Aggregate-based map/reduce program that counts the words in the input files.\\ \hline                  
            2&  aggregatewordhist  &An Aggregate-based map/reduce program that computes the histogram of the words in the input files. \\ \hline      
            3&  bbp  &A map/reduce program that uses Bailey-Borwein-Plouffe to compute exact digits of Pi.\\ \hline
            4&  distbbp  &A map/reduce program that uses a BBP-type formula to compute exact bits of Pi. \\ \hline
            5&  grep  &A map/reduce program that counts the matches of a regex in the input.\\ \hline
            6&  pi  &A map/reduce program that estimates Pi using a quasi-Monte Carlo method.\\ \hline              
            7&  randomtextwriter  &A map/reduce program that writes 10GB of random textual data per node.\\ \hline              
            8&  randomwriter  &A map/reduce program that writes 10GB of random data per node.\\ \hline              
            9&  sort  &A map/reduce program that sorts the data written by the random writer.\\ \hline
            10&  teragen  &Generate data for the terasort.\\ \hline
            11&  terasort  &Run the terasort.\\ \hline
            12& teravalidate  &Check the results of the terasort.\\ \hline               
            13&  wordcount &A map/reduce program that counts the words in the input files.\\ \hline
            14&  wordmean  &A map/reduce program that counts the average length of the words in the input files.\\ \hline
            15&  wordmedian  &A map/reduce program that counts the median length of the words in the input files.\\ \hline
            16&  wordstandarddeviation  &A map/reduce program that counts the standard deviation of the length of the words in the input files.\\ \hline
      \end{tabulary}
\end{table}

\begin{table}
      \caption{List of Spark-Perf MLlib Tests}
      \label{table_examples_spark}
      \centering
      \begin{tabulary}{0.45\textwidth}{| C | C | C |}\hline
      \textbf{Exp.no}& \textbf{Name} &\textbf{Description}\\ \hline
      1&fp-growth&  Frequent Pattern Matching Tests to find frequent item sets \\   \hline
      2&word2vec&  Feature Transformation Tests for distributed presentation of words  \\   \hline
      3&chi-sq-feature&  Statistic Toolkit Tests using Chi-square for correlation \\   \hline
      4&spearman&  Statistic Toolkit Tests using Spearman's Correlation  \\   \hline
      5&pearson &  Statistic Toolkit Tests using Pearson's Correlation    \\   \hline
      6&block-matrix-mult&  Matrix Multiplication on distributed matrix\\   \hline
      7&summary-statistics&  Linear Algebra Tests using Summary Statistics (min, max, ...)   \\   \hline
      8&pca& Linear Algebra Tests using Principal Component Analysis   \\   \hline
      9&svd& Linear Algebra Tests using Singular Value Decomposition   \\   \hline
      10&gmm& Clustering Tests using Gaussian Mixture Model  \\   \hline
      11&kmeans& Clustering Tests using K-Means clustering   \\   \hline
      12&als& Recommendation Tests using Alternating Least Squares   \\   \hline
      13&decision-tree& Random Forest Decision Tree \\   \hline
      14&naive-bayes& Classification Tests using Naive Bayes   \\   \hline
      15&glm-classification&  Generalized Linear Classification Model    \\   \hline
      16&glm-regression&  Generalized Linear Regression Model    \\   \hline
      \end{tabulary}
\end{table}


Though the main requirement for any attack detection service is to be able to detect an attack successfully, being able to detect the attack before the attacked program completes execution is also a necessity. We show the efficiency and the overhead of the proposed system by conducting the experiments in real-time using popular examples and tests. We used two sets of open-source big data benchmark programs: (a) 16 \textit{Hadoop MapReduce Examples:} that are provided in the Apache hadoop installation kit; and (b) 16 \textit{Spark-perf MLlib Tests:} for machine learning algorithms given in the spark performance test suite by Databricks \cite{Databricks}. More details about these examples and tests are given in tables \ref{table_examples_hadoop} and \ref{table_examples_spark}.  

The input to our model (built from the proposed system) is the run-time assembly code of a program. The hadoop mapreduce examples were coded in Java and the Spark-perf MLlib tests were coded in Scala. So, the jars to run these examples were built using just-in-time compiling. Their bytecodes are insufficient to create the assembly codes of the individual programs. We used a software called jit-watch \cite{Jitwatch} to generate the assembly codes (Intel x86 specification) of the programs from the jars. Since our algorithm only needs control-flow instructions from the generated assembly code outputs of each program, we used a custom parser that can filter out control flow instructions from the native files. All 32 example programs are infected by a code snippet that calls a function \texttt{foo} to print a line to the console and involves a total of 3 \texttt{call} instructions and 1 \texttt{return} instruction. The command used for generating assembly code output of JVM (or Hotspot VM) when running the program is: \newline
\texttt{java -XX:+UnlockDiagnosticVMOptions -XX:+PrintAssembly -XX:PrintAssemblyOptions=intel -XX:+TraceClassLoading -XX:+LogCompilation -XX:LogFile=filename -cp [path to additional classes] [main method] [args]} 

First, we calculated the execution times for the hadoop mapreduce examples on the hadoop cluster. Then we studied the run times of the implemented model while it was analyzing the assembly codes of the driver programs of the same examples. These experiments are adhoc because the input arguments for some of the experiments were intentionally low to simulate worst case scenario's where the process takes very less time to execute. To meet the input data requirements of the mapreduce examples, we put the configuration file data from \texttt{etc} folder of hadoop into HDFS. The generic command used to run these mapreduce examples is: \newline
\texttt{time hadoop jar hadoop-mapreduce-examples.jar [main method] [args]}

\begin{table*}[!t]
  \caption{Hadoop Map Reduce Examples - Instruction level properties}
  \label{table_hadoop_values}
  \centering
  \begin{tabular}{|c|c|c|c|c|c|c|c|c|c|c|}
      \hline
    \textbf{Exp.no}& \textbf{Example}&  \textbf{Instruction Count}  &       \textbf{CFI}&   \textbf{Jumps}& \textbf{Calls}  & \textbf{Returns}& \textbf{\%CFI} &  \textbf{\% Jumps} & \textbf{\% Calls} & \textbf{\% Returns}\\
    
    \hline
1&  aggregatewordcount& 81713&  17195&  12722&  4009& 464&  21.04\%&  15.57\%&  4.91\%& 0.57\%  \\\hline
2&  aggregatewordhist&  48428&  9812& 7133& 2366& 313&  20.26\%&  14.73\%&  4.89\%& 0.65\%  \\\hline
3&  bbp&  85514&  17880&  13182&  4211& 487&  20.91\%&  15.42\%&  4.92\%& 0.57\%  \\\hline
4&  distbbp&  68283&  13880&  10234&  3238& 408&  20.33\%&  14.99\%&  4.74\%& 0.60\%  \\\hline
5&  grep& 81404&  16911&  12501&  3937& 473&  20.77\%&  15.36\%&  4.84\%& 0.58\%  \\\hline
6&  pi& 65397&  13607&  10170&  3070& 367&  20.81\%&  15.55\%&  4.69\%& 0.56\%  \\\hline
7&  randomtextwriter& 70909&  14896&  11186&  3332& 378&  21.01\%&  15.78\%&  4.70\%& 0.53\%  \\\hline
8&  randomwriter& 91414&  19462&  14508&  4475& 479&  21.29\%&  15.87\%&  4.90\%& 0.52\%  \\\hline
9&  sort& 101298& 21420&  16003&  4885& 532&  21.15\%&  15.80\%&  4.82\%& 0.53\%  \\\hline
10& teragen&  134747& 28228&  21013&  6516& 699&  20.95\%&  15.59\%&  4.84\%& 0.52\%  \\\hline
11& terasort& 121541& 25420&  18925&  5827& 668&  20.91\%&  15.57\%&  4.79\%& 0.55\%  \\\hline
12& teravalidate& 139583& 29244&  21838&  6630& 776&  20.95\%&  15.65\%&  4.75\%& 0.56\%  \\\hline
13& wordcount&  77393&  16341&  12100&  3791& 450&  21.11\%&  15.63\%&  4.90\%& 0.58\%  \\\hline
14& wordmean& 62412&  13093&  9726& 2994& 373&  20.98\%&  15.58\%&  4.80\%& 0.60\%  \\\hline
15& wordmedian& 66401&  13435&  9869& 3161& 405&  20.23\%&  14.86\%&  4.76\%& 0.61\%  \\\hline
16& wordstandarddeviation&  82079&  16917&  12492&  3932& 493&  20.61\%&  15.22\%&  4.79\%& 0.60\%  \\\hline
\multicolumn{2}{|c|}{Average Values}& 86157&  17984&  13350&  4148& 485&  \textbf{20.83\%}& \textbf{15.45\%}& \textbf{4.81\%}&  \textbf{0.57\%} \\\hline
  \end{tabular}
\end{table*}

\begin{table*}[!t]
  \caption{Spark Performance Test ML Algorithms - Instruction level properties}
  \label{table_spark_values}
  \centering
  \begin{tabular}{|c|c|c|c|c|c|c|c|c|c|c|}
      \hline
    \textbf{Exp.no}& \textbf{Algorithm}&  \textbf{Instruction Count}  &       \textbf{CFI}&   \textbf{Jumps}& \textbf{Calls}  & \textbf{Returns}& \textbf{\% CFI} & \textbf{\%  Jumps}  & \textbf{\% Calls} & \textbf{\% Returns}\\
    \hline
1&  fp-growth&  216009& 46544&  35200&  10251&  1093& 21.55\%&  16.30\%&  4.75\%& 0.51\%  \\\hline
2&  word2vec& 147737& 30235&  22638&  6772& 825&  20.47\%&  15.32\%&  4.58\%& 0.56\%  \\\hline
3&  chi-sq-feature& 172014& 35783&  26736&  8119& 928&  20.80\%&  15.54\%&  4.72\%& 0.54\%  \\\hline
4&  spearman& 194615& 41043&  30857&  9155& 1031& 21.09\%&  15.86\%&  4.70\%& 0.53\%  \\\hline
5&  pearson&  184628& 38694&  28996&  8691& 1007& 20.96\%&  15.71\%&  4.71\%& 0.55\%  \\\hline
6&  block-matrix-mult&  195714& 41245&  31030&  9174& 1041& 21.07\%&  15.85\%&  4.69\%& 0.53\%  \\\hline
7&  summary-statistics& 196555& 41034&  30736&  9235& 1063& 20.88\%&  15.64\%&  4.70\%& 0.54\%  \\\hline
8&  pca&  192280& 40427&  30377&  9020& 1030& 21.03\%&  15.80\%&  4.69\%& 0.54\%  \\\hline
9&  svd&  143996& 29684&  22334&  6550& 800&  20.61\%&  15.51\%&  4.55\%& 0.56\%  \\\hline
10& gmm&  170722& 35655&  26848&  7898& 909&  20.88\%&  15.73\%&  4.63\%& 0.53\%  \\\hline
11& kmeans& 170694& 35842&  26957&  7962& 923&  21.00\%&  15.79\%&  4.66\%& 0.54\%  \\\hline
12& als&  181836& 38032&  28603&  8428& 1001& 20.92\%&  15.73\%&  4.63\%& 0.55\%  \\\hline
13& decision-tree&  175889& 36655&  27546&  8140& 969&  20.84\%&  15.66\%&  4.63\%& 0.55\%  \\\hline
14& naive-bayes&  171945& 36053&  27036&  8082& 935&  20.97\%&  15.72\%&  4.70\%& 0.54\%  \\\hline
15& glm-classification& 186454& 39088&  29362&  8715& 1011& 20.96\%&  15.75\%&  4.67\%& 0.54\%  \\\hline
16& glm-regression& 200255& 42439&  32020&  9346& 1073& 21.19\%&  15.99\%&  4.67\%& 0.54\%  \\\hline
\multicolumn{2}{|c|}{Average Values} &  181334& 38028&  28580&  8471& 977&  \textbf{20.95\%}& \textbf{15.74\%}& \textbf{4.67\%}&  \textbf{0.54\%} \\\hline

  \end{tabular}
\end{table*}

\begin{figure*}[!ht]
  \centering
    \subfloat[Hadoop Map-Reduce examples\label{subfig_hadoop}]{%
      \includegraphics[width=0.47\textwidth]{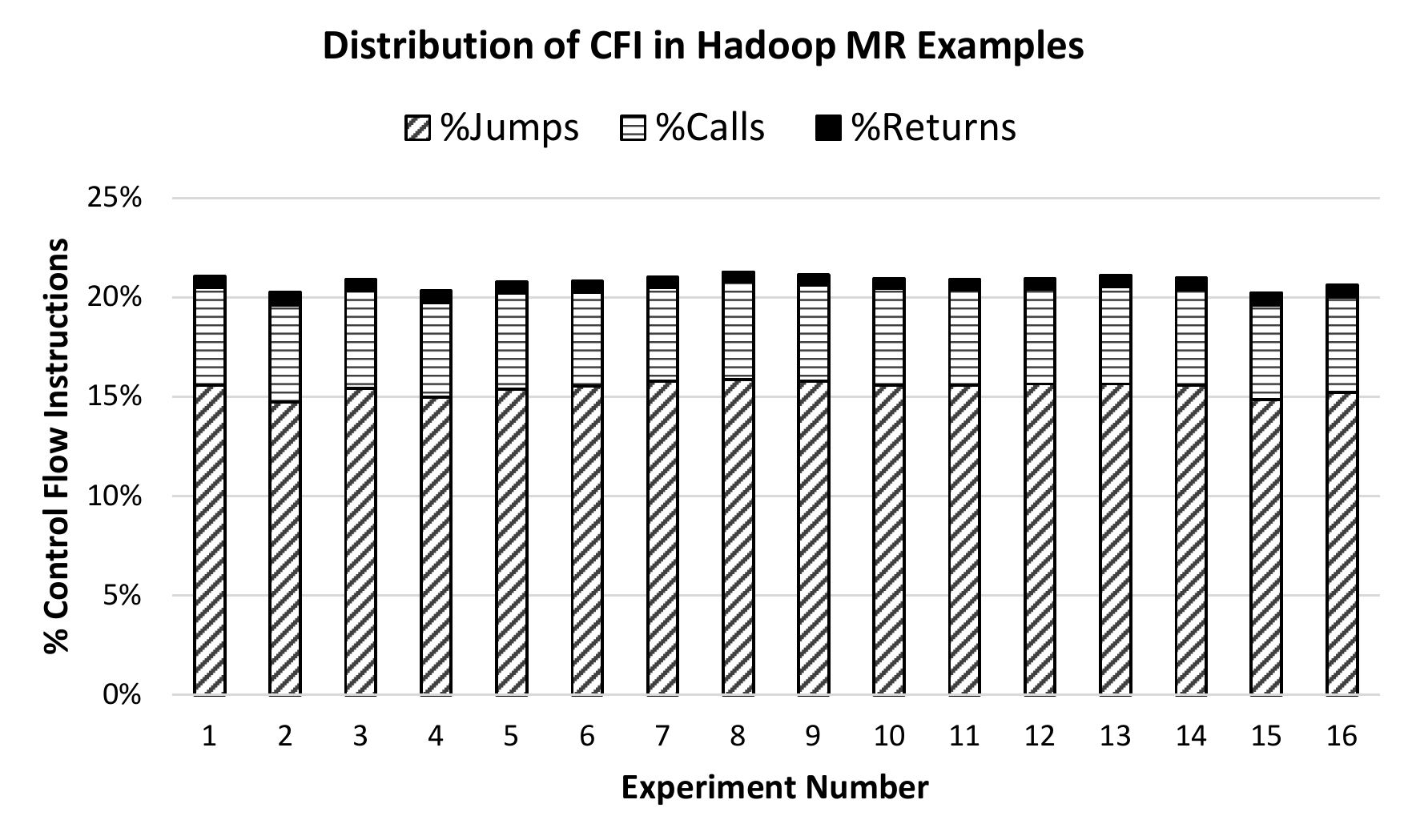}
    }
    \hfill
    \subfloat[Spark-Perf Machine Learning Tests\label{subfig_spark}]{%
      \includegraphics[width=0.47\textwidth]{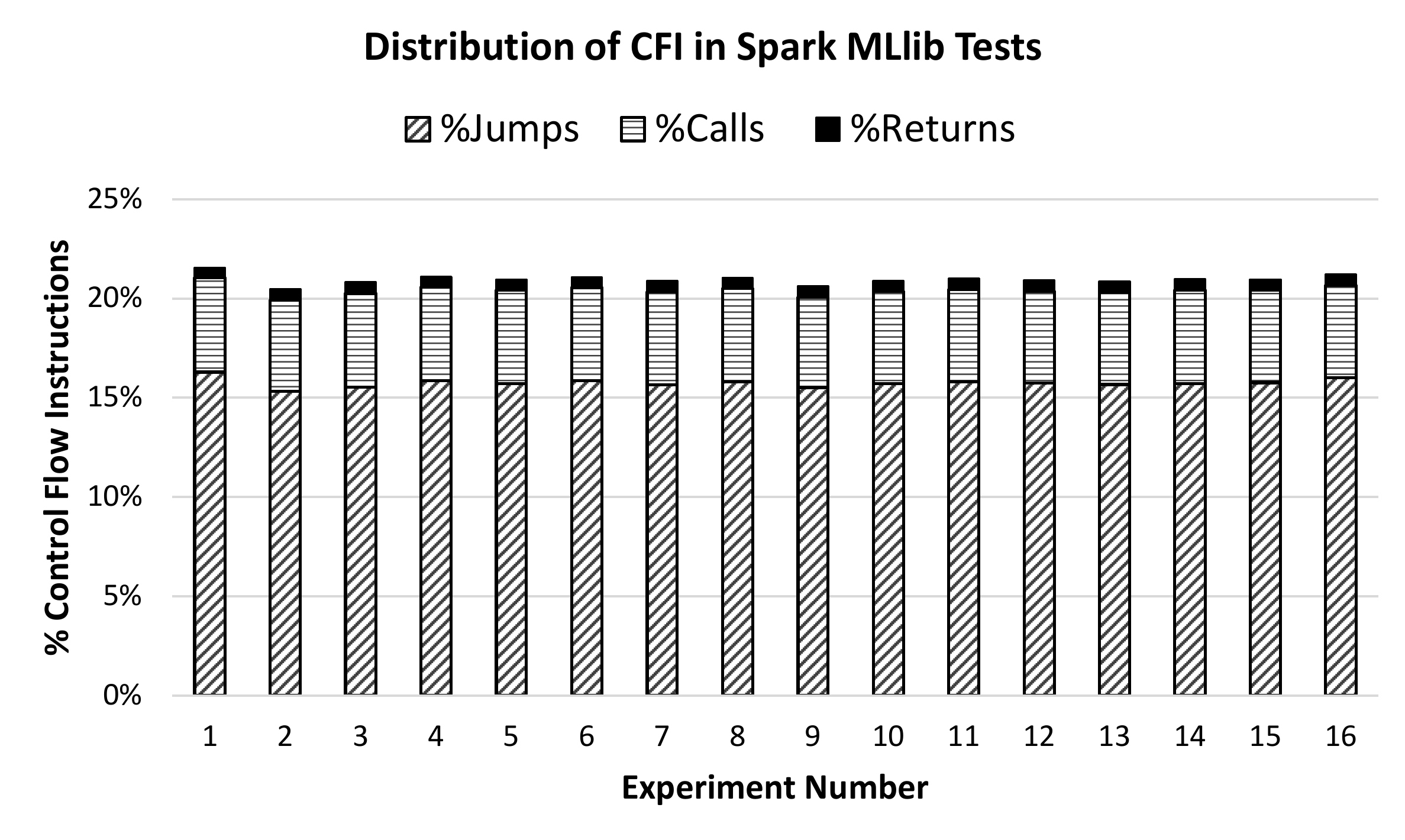}
    }
    \caption{Consistent Distribution of CFI in the Programs used for Experimental Evaluation}
    \label{fig_distcfi}
\end{figure*}

The spark-perf MLlib tests on the spark cluster were conducted the same way the mapreduce examples were tested. But here the inputs for the tests were predetermined by the benchmark provider in the \texttt{config.py} script. The generic command used to run these MLlib tests is:  \newline
\texttt{spark-submit --class mllib.perf.TestRunner --master [ip of node] --driver-memory [limit] mllib-perf-tests-assembly.jar [algorithm name] [args]}

\subsection{Results and Analysis}

\begin{figure*}[!ht]
    \subfloat[Run Time Results for Hadoop MapReduce Examples \label{subfig_time_hadoop}]{%
      \includegraphics[width=0.48\textwidth, height=2.35in]{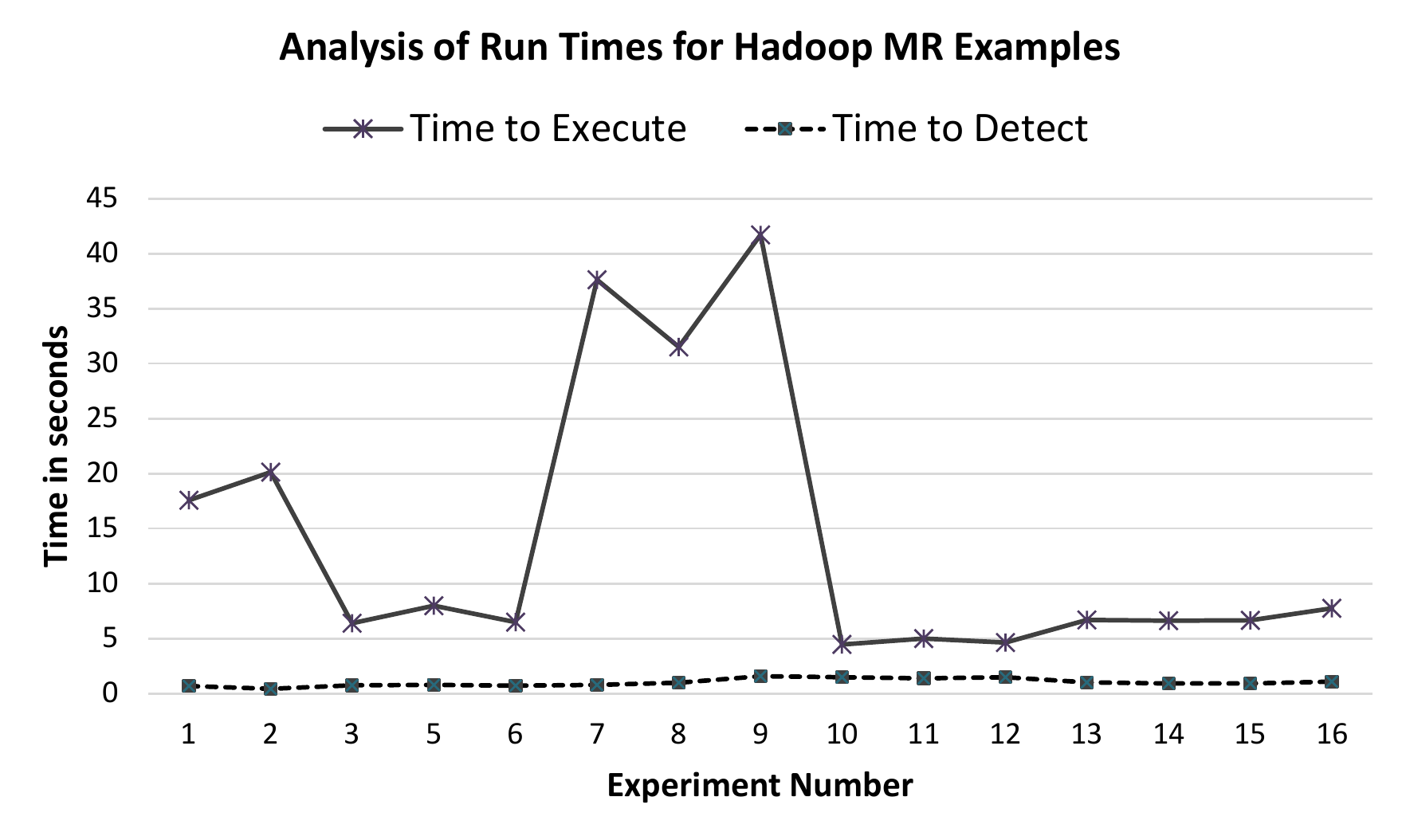}
    }
    \hfill
    \subfloat[Run Time Results for Spark-perf MLlib Tests \label{subfig_time_spark}]{%
      \includegraphics[width=0.48\textwidth, height=2.35in]{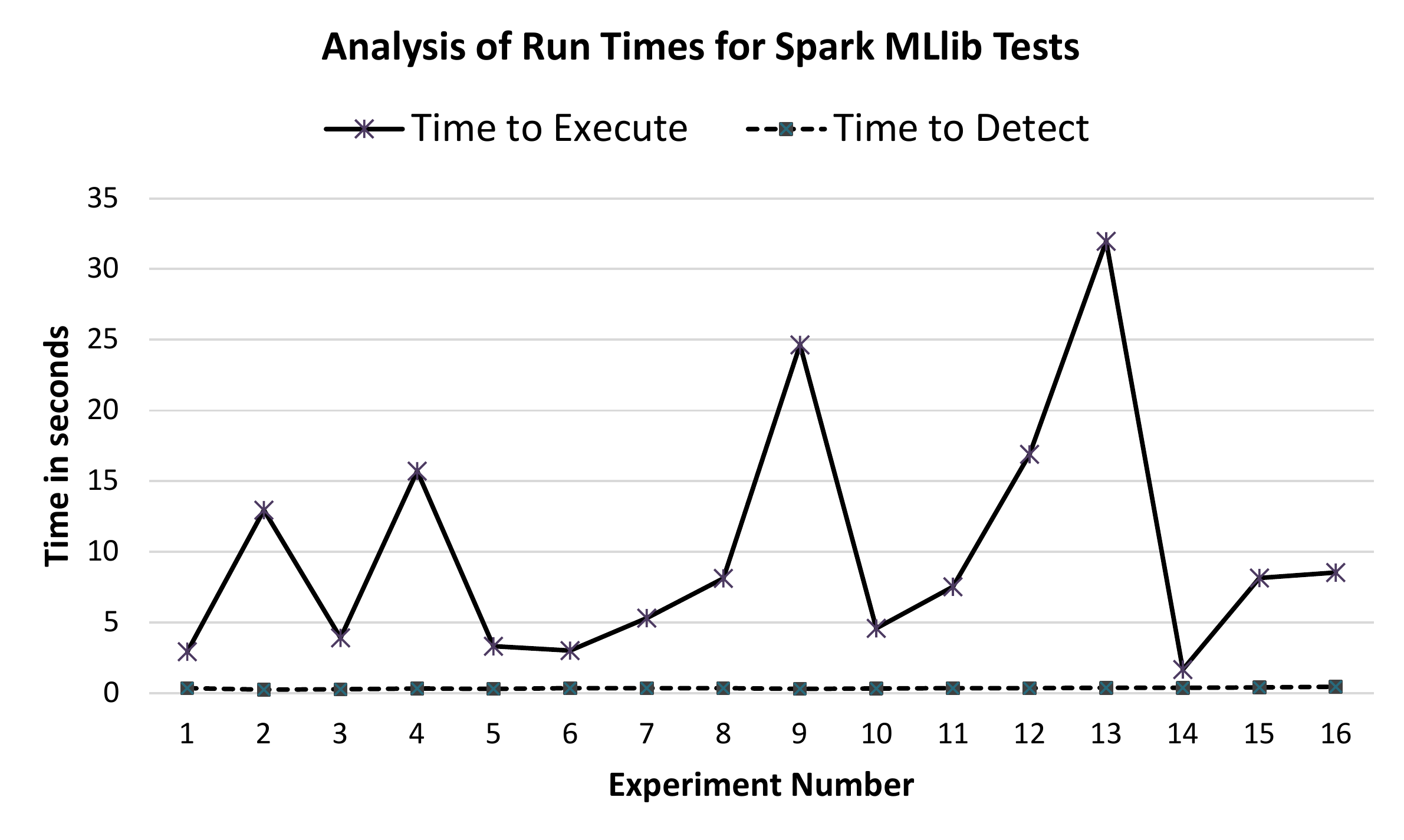}
    }
    \caption{Run-time Analysis of Hadoop and Spark programs}
    \label{fig_time_comparison}
\end{figure*}

\begin{figure*}[!ht]
    \subfloat[Time Forecast Plot for Detection Algorithm on Hadoop Cluster \label{subfig_forecast_hadoop}]{%
      \includegraphics[width=0.48\textwidth, height=2.35in]{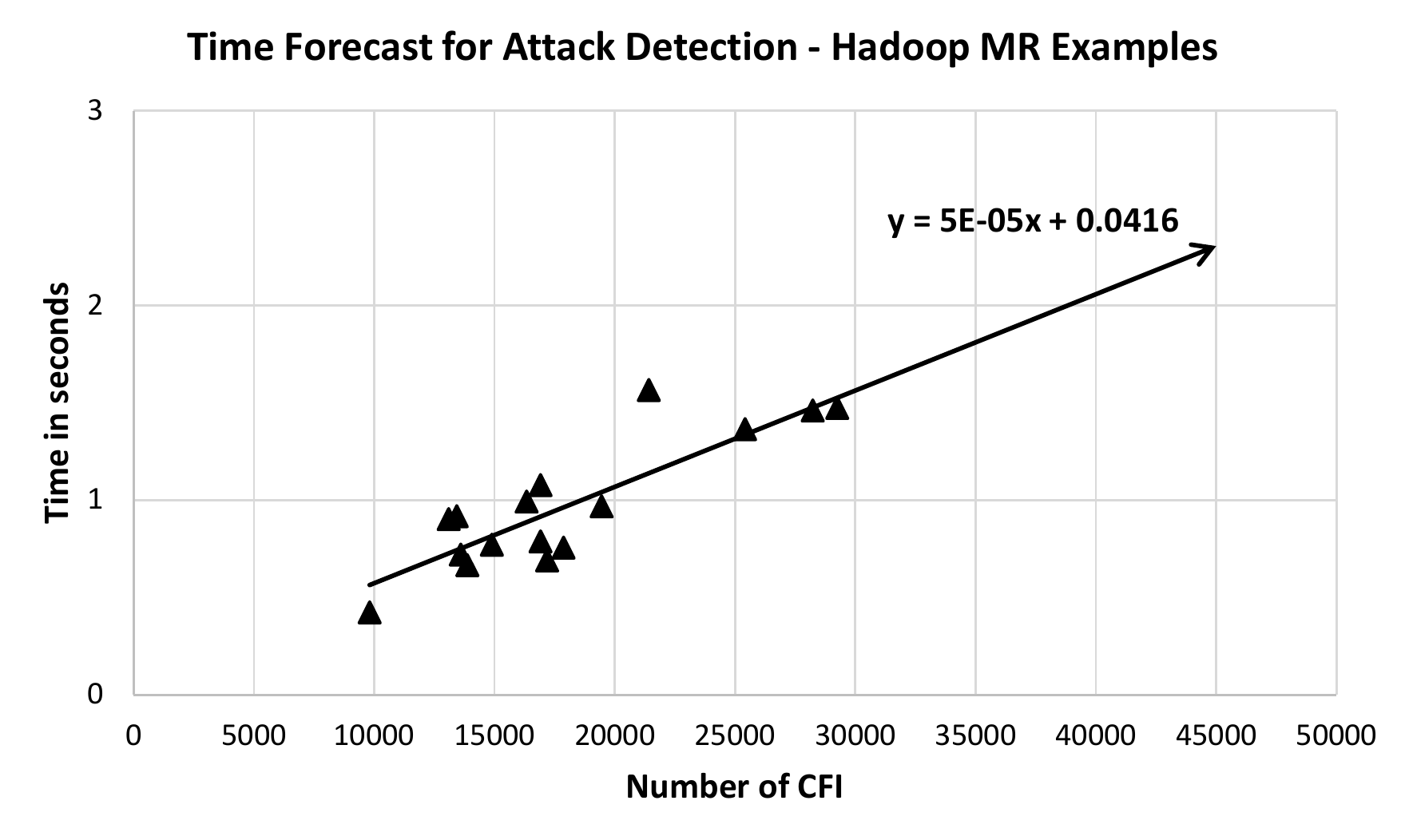}
    }
    \hfill
    \subfloat[Time Forecast Plot for for Detection Algorithm on Spark Cluster \label{subfig_forecast_spark}]{%
      \includegraphics[width=0.48\textwidth, height=2.35in]{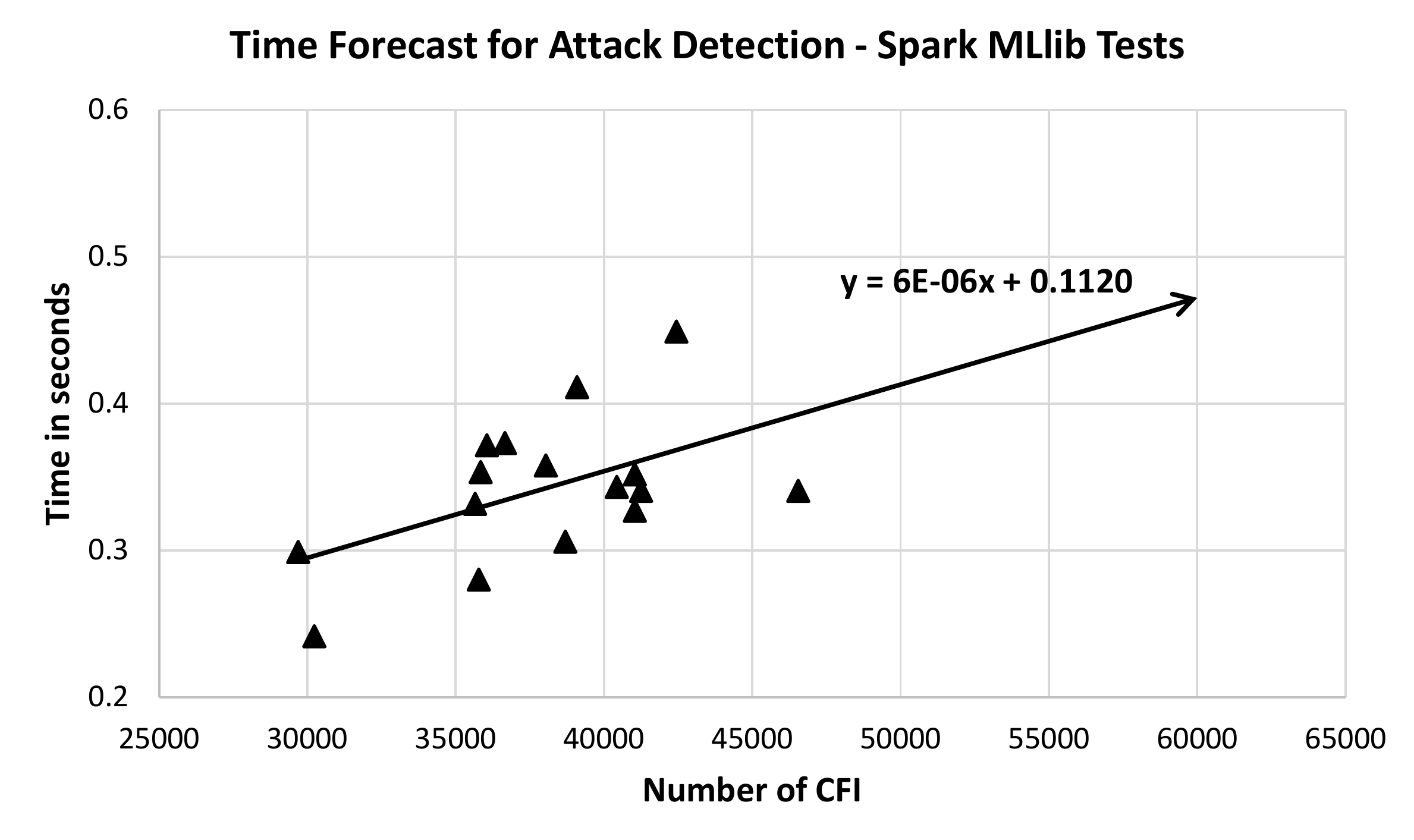}
    }
    \caption{Time Forecast Plots (best-fit regression) of the Model built from Proposed System}
    \label{fig_time}
\end{figure*}

The experiments we used for evaluating our proposed security system comprise of stress tests and performance benchmarks of hadoop and spark. Hence, knowing which threads of investigation to follow and which to ignore was difficult and challenging. We chose to focus on execution time and code size of the experiments. The overhead in our experiments is calculated from time measurements. We divide the time taken to detect an attack in a process $p$ by the execution time of the same process and multiply the result by 100 to find the percentage of time overhead, as given in equation \ref{eq:overhead}. Here $time_{detect}(p)$ is calculated using system clock measurements for encrypting process analysis information, decrypting received messages and hash matching. The communication cost in sending data packets from one node to another is not included. The overhead calculations show the worst case scenario since the input arguments are intentionally low for some of the experiments. Real-world big data programs will be much more complex jobs and hence the overhead will be much lesser than what is shown here. Tables 6 and 7 and Figures \ref{subfig_time_hadoop} and \ref{subfig_time_spark} show the analysis of run-times for executing the experiments and the model built from the proposed system. On average, the overhead of running the model is 3.28\%. We used linear regression and best-fit plots, given in Figures \ref{subfig_forecast_hadoop} and \ref{subfig_forecast_spark}, to show the relation between programs (given in number of control flow instructions of their assembly representations) and time to detect an attack in them. The time taken to execute example number 4 i.e.\ \textit{distributed bbp} program of hadoop mapreduce example set was too high (288 seconds) to plot on the graph shown in Figure \ref{subfig_time_hadoop}.   

%

\begin{equation}
\label{eq:overhead}
\% overhead(p) = \frac{time_{detect}(p)}{time_{execute}(p)} \times 100
\end{equation}


\begin{table}
    \label{table_hadoop_time}
      \caption{Run Time Analysis for Proposed system to analyze and compare Hadoop MapReduce Examples}
      \centering
      \begin{tabulary}{0.45\textwidth}{|C|C|C|C|}
      \hline
      \textbf{Exp.no} & \textbf{Time to Execute} & \textbf{Time to Detect} & \textbf{\% Overhead}\\ \hline      
        1&    17.56&  0.69& 3.93\%  \\\hline
        2&    20.14&  0.42& 2.10\%  \\\hline
        3&    6.39& 0.76& 11.84\% \\\hline
        4&    287.62& 0.67& 0.23\%  \\\hline
        5&    7.96& 0.79& 9.89\%  \\\hline
        6&    6.48& 0.72& 11.12\% \\\hline
        7&    37.63&  0.77& 2.05\%  \\\hline
        8&    31.51&  0.97& 3.07\%  \\\hline
        9&    41.71&  1.57& 3.75\%  \\\hline
        10&   4.45& 1.46& 32.82\% \\\hline
        11&   4.99& 1.37& 27.37\% \\\hline
        12&   4.61& 1.47& 31.96\% \\\hline
        13&   6.68& 0.99& 14.86\% \\\hline
        14&   6.63& 0.90& 13.63\% \\\hline
        15&   6.64& 0.92& 13.82\% \\\hline
        16&   7.76& 1.08& 13.88\% \\\hline
        Average Values& 31.17&  0.97& \textbf{3.12\%} \\\hline
      \end{tabulary}
\end{table}      

\begin{table}
    \label{table_spark_time}
      \caption{Run Time Analysis for Proposed system to analyze and compare Spark-perf MLlib Tests}
      \centering
      \begin{tabulary}{0.45\textwidth}{|C|C|C|C|}
      \hline
      \textbf{Exp.no} & \textbf{Time to Execute} & \textbf{Time to Detect} & \textbf{\% Overhead}\\ \hline  
        1&    2.92& 0.34& 11.67\% \\\hline
        2&    12.942& 0.24& 1.87\%  \\\hline
        3&    3.899&  0.28& 7.19\%  \\\hline
        4&    15.708& 0.33& 2.08\%  \\\hline
        5&    3.314&  0.31& 9.23\%  \\\hline
        6&    3.011&  0.34& 11.31\% \\\hline
        7&    5.312&  0.35& 6.63\%  \\\hline
        8&    8.124&  0.34& 4.23\%  \\\hline
        9&    24.647& 0.30& 1.21\%  \\\hline
        10&   4.584&  0.33& 7.24\%  \\\hline
        11&   7.529&  0.35& 4.69\%  \\\hline
        12&   16.884& 0.36& 2.12\%  \\\hline
        13&   31.963& 0.37& 1.17\%  \\\hline
        14&   1.664&  0.37& 22.34\% \\\hline
        15&   8.151&  0.41& 5.05\%  \\\hline
        16&   8.542&  0.45& 5.26\%  \\\hline
        Average Values& 9.950&  0.34& \textbf{3.44\%} \\\hline
      \end{tabulary}
\end{table}

The proposed system performs a similarity check of control flow within duplicate processes running on different nodes of a big data cluster. This control flow similarity check is performed by matching control instruction sequences. Since the infected node is predetermined in our experiments, our test cases do not have a false positive or false negative. But a false positive will occur when all data nodes are attacked in the same way. A false negative will occur in case of runtime attacks or attacks that originate outside the big data platform. But given our attack model, such cases are outside the scope of this work. Instead, we try to understand the control flow in the programs used in the experiments section, i.e.\ hadoop mapreduce examples and the spark performance tests for machine learning algorithms. Results from tables \ref{table_hadoop_values} and \ref{table_spark_values} and fugures \ref{subfig_hadoop} and \ref{subfig_spark} show instruction level properties of the examples and tests used in our experiments. It can be observed that only 20.8\% of the total instruction count in the hadoop mapreduce examples account for control flow instructions. In case of spark performance tests for machine learning algorithms, 20.9\% of instructions in the assembly code are control flow instructions. Of all control flow instructions, \texttt{jumps} are the most significantly used CFI with a lion share of 15.45\% of the total instruction count in hadoop mapreduce examples and 15.74\% of the total instruction count in spark performance tests. \texttt{Calls} and \texttt{returns} cover only 4.8\% and 0.5\% respectively in the hadoop mapreduce example set and; 4.6\% and 0.5\% respectively in the spark performance tests set. 

It can be inferred from these results that control flow instructions account for only one-fifth of the total instruction count for a program (assembly code). This is a remarkable coincidence among these two sets of programs because (a) they belong to different domains - mapreduce on hadoop, machine learning in spark; (b) their source programming language is different - java for hadoop mapreduce examples, scala for spark-perf machine learning tests; and (c) they differ in program size - 86,000 instructions on average per program for the mapreduce example set and 180,000 instructions on average per program for the spark perf machine learning tests. This observation  strengthens our initial argument that generating dynamic CFG for large and complex big data programs is cumbersome. This is because the size of CFG is proportional to the code lines which is related to the number of instructions. Hence, the proposed idea of generating CIS and hashing them is a good alternative to the CFG memory complexity problem. The overhead incurred in using the model built from the proposed system architecture is less than 3.28\% if it is hosted by the same hardware that hosts the big data systems. This is in the acceptable range of overhead for big data platforms like Hadoop. The time our system takes to analyze the programs and compare the results is linearly dependent on the number of control flow instructions in the program, but not on the number of lines of assembly code. This greatly reduces the complexity of the similarity analysis from the conventional and complex approach of generating a CFG. Also, generating CIS only needs a one time parse through the program code (assembly code) and can be performed independently and in parallel on each node of the cluster. The experimental results show the feasibility of implementing a model of the proposed system. Building and implementing a detailed version of this system will demonstrate lower overhead and convince the vendors to adopt it. 

\section{Conclusion}\label{sec:conclusion}
In this paper, we proposed a security system for big data systems to detect insider attacks quickly with low overhead. The system consists of a two step attack detection algorithm and a secure communication protocol. A simple hash string matching technique is proposed to fulfill the distributed process similarity check and identify attacks. A secure communication protocol for data nodes that uses periodically generated random keys is proposed to conduct the detection algorithm. A model of the proposed system is tested in real-time on Amazon's EC2 clusters using a different sets of Hadoop and Spark programs. The time overhead was 3.28\% and it is observed from the results that the proposed security system uses only 20\% of program code to detect attacks. In this work, we also propose the idea of delegating security as an independent module and the components needed for such models are discussed. For future work, we would like to evaluate our system on security related big data benchmarks (when available). Also, we would like to actualize the hardware architecture of security chips that can independently support our system.


%

\begin{IEEEbiographynophoto}{Santosh Aditham} received the B.Tech. degree in computer science from Andhra University, Visakhapatnam, India, in 2007 and the M.S. degree in computer science  from the Texas Tech University, Lubbock, TX in 2010. Currently he is a PhD candidate in computer science and engineering department from the University of South Florida, Tampa, FL. He likes to work at computer architecture and operating systems level. His research interests include security of big data systems and low-power scheduling in distributed systems. Mr. Aditham is a student member of the IEEE Computer Society.
\end{IEEEbiographynophoto}
\begin{IEEEbiographynophoto}{Nagarajan "Ranga" Ranganathan} received the B.E. (Honors) degree in Electrical and Electronics Engineering from Regional Engineering College (National Institute of Technology) Tiruchi, University of Madras, India, 1983, and the Ph.D. degree in Computer Science from the University of Central Florida, Orlando in 1988. He is a Distinguished University Professor Emeritus of Computer Science and Engineering at the University of South Florida, Tampa. During 1998-99, he was a Professor of Electrical and Computer Engineering at the University of Texas at El Paso. His research interests include VLSI circuit and system design, VLSI design automation, multi-metric optimization in hardware and software systems, computer architecture, reversible logic and parallel computing. He has developed many special purpose VLSI circuits and systems for computer vision, image and video processing, pattern recognition, data compression and signal processing applications. He and his students have developed several VLSI CAD algorithms based on decision theory, game theory, auction theory and Fuzzy modeling. He has co-authored about 300 papers in refereed journals and conferences, five book chapters and co-owns eight U.S. patents and two pending. Dr. Ranganathan was elected as a Fellow of IEEE in 2002 for his contributions to algorithms and architectures for VLSI systems and as Fellow of AAAS in 2012.
\end{IEEEbiographynophoto}




\end{document}